\documentclass[a4paper,aps,prd,twocolumn,showpacs]{revtex4-1}
\pdfoutput=1

\usepackage{graphicx}
\usepackage[caption=false,farskip=0pt]{subfig}
\usepackage{hyperref}

\usepackage{bm}
\usepackage{amsmath}
\usepackage{amssymb}

\newcommand{\Ans}[1][n]{A_{#1}^{*}}
\newcommand{\Zn}[1][n]{\bbZ^{#1}}
\newcommand{\avg}[1]{\langle{#1}\rangle}
\newcommand{\bbR}{\mathbb{R}}
\newcommand{\bbZ}{\mathbb{Z}}
\newcommand{\calA}{\mathcal{A}}
\newcommand{\calF}{\mathcal{F}}
\newcommand{\calN}{\mathcal{N}}
\newcommand{\calP}{\mathcal{P}}
\newcommand{\inv}[1][1]{^{-#1}}

\newcommand{\mat}[1]{\mathbf{#1}}
\newcommand{\ndot}[2][s]{#2^{(#1)}}
\newcommand{\pr}{^{\,\prime}}
\newcommand{\relerr}[2]{\varepsilon(#1,#2)}
\newcommand{\trsp}{^{\mathrm{T}}}

\newcommand{\ua}{_{a}}
\newcommand{\ub}{_{b}}

\newcommand{\uest}{_{\mathrm{est}}}
\newcommand{\ufreq}{_{\mathrm{freq}}}
\newcommand{\umax}{_{\mathrm{max}}}

\newcommand{\unana}{_{n\ua n\ua}}
\newcommand{\unbnb}{_{n\ub n\ub}}

\newcommand{\uref}{_{0}}
\newcommand{\urssnn}{_{\mathrm{rss},\vec n \vec n}}
\newcommand{\urssnunu}{_{\mathrm{rss},\ndot\nu \ndot\nu}}
\newcommand{\urss}{_{\mathrm{rss}}}
\newcommand{\usky}{_{\mathrm{sky}}}
\newcommand{\ussfn}{_{\mathrm{ss},\ndot f \vec n}}
\newcommand{\ussnf}{_{\mathrm{ss},\vec n \ndot f}}
\newcommand{\ussnn}{_{\mathrm{ss},\vec n \vec n}}
\newcommand{\uss}{_{\mathrm{ss}}}
\newcommand{\ustart}{_{\mathrm{start}}}
\newcommand{\utwoF}{_{0}}
\newcommand{\ux}{_{x}}
\newcommand{\uy}{_{y}}
\newcommand{\uz}{_{z}}

\newcommand{\Ulatt}[1][]{^{\;\mathrm{\ell}#1}}
\newcommand{\Umax}{^{\mathrm{max}}}
\newcommand{\Umin}{^{\mathrm{min}}}
\newcommand{\Usig}{^{\mathrm{s}}}
\newcommand{\Utmpl}{^{\mathrm{t}}}

\newcommand{\commitDATE}{2014-12-22 09:50:44 +0100}

\newcommand{\commitIDshort}{commitID: cb5ec5f}
\newcommand{\commitSTATUS}{CLEAN}

\defcitealias{Wette.Prix.2013a}{Paper~I}
\newcommand{\PaperI}{\citetalias{Wette.Prix.2013a}}

\begin{document}

\title{Lattice template placement for coherent all-sky searches for gravitational-wave pulsars}
\author{Karl Wette}
\email{karl.wette@aei.mpg.de}
\affiliation{Max-Planck-Institut f\"ur Gravitationsphysik (Albert-Einstein-Institut), Callinstr.\ 38, 30167 Hannover, Germany}

\date{\commitDATE; \commitIDshort-\commitSTATUS}

\begin{abstract}
All-sky, broadband, coherent searches for gravitational-wave pulsars are restricted by limited computational resources.
Minimizing the number of templates required to cover the search parameter space, of sky position and frequency evolution, is one important way to reduce the computational cost of a search.
We demonstrate a practical algorithm which, for the first time, achieves template placement with a minimal number of templates for an all-sky search, using the reduced supersky parameter-space metric of Wette and Prix [Phys.~Rev.~D \textbf{88}, 123005 (2013)].
The metric prescribes a constant template density in the signal parameters, which permits that templates be placed at the vertices of a lattice.
We demonstrate how to ensure complete coverage of the parameter space, including in particular at its boundaries.
The number of templates generated by the algorithm is compared to theoretical estimates, and to previous predictions by Brady \emph{et~al.} [Phys.~Rev.~D \textbf{57}, 2101 (1998)].
The algorithm may be applied to any search parameter space with a constant template density, which includes semicoherent searches and searches targeting known low-mass X-ray binaries.
\end{abstract}

\pacs{04.80.Nn, 95.55.Ym, 95.75.Pq, 97.60.Jd}

\maketitle

\section{Introduction}\label{sec:introduction}

Gravitational-wave pulsars are rapidly-spinning neutron stars which could be emitting gravitational radiation if nonaxisymmetrically deformed, due to various mechanisms which may support such a deformation (see~\cite{Prix.2009a,Sathya.Schutz.2009} for reviews of emission mechanisms, and~\cite{Owen.2005a,JohnsonMcDaniel.Owen.2013a} for predictions of the maximum supportable nonaxisymmetry).
Their detection by ground-based interferometric detectors with kilometer-long arms, such as LIGO~\cite{Abbott.etal.2009f} and Virgo~\cite{Accadia.etal.2012a}, is one of the great challenges of gravitational-wave physics.
Searches for gravitational-wave pulsars in data from the first generation of interferometric detectors~\cite[e.g.][]{Abadie.etal.2012b,Aasi.etal.2013a,Aasi.etal.2013c} have, to date, not yielded a detection.
Indeed, gravitational-wave pulsars are expected to be difficult to detect even by the next generation of detectors~\cite{Harry.etal.2010a,Aasi.etal.2013d,Somiya.2012a}, which are currently under construction.
Maximizing the chance of a detection requires investment in highly-optimized data analysis techniques, and large-scale computing resources such as Einstein@Home~\cite{Aasi.etal.2013a}.

The signals emitted by gravitational-wave pulsars are characteristically continuous, narrow-band, and quasi-sinusoidal, and are believed to be well-modeled by a parameterized template waveform family.
The most sensitive search method is thus coherent matched filtering against a bank of templates, whose parameters are chosen from a space of interest; this yields the detection statistic commonly known in the field as the $\calF$-statistic~\cite{Jaranowski.etal.1998a,Cutler.Schutz.2005a}.
This method is, however, too computationally intensive to search year-long data sets and wide parameter spaces, such as all-sky broadband-frequency searches for undiscovered gravitational-wave pulsars.
For such searches, a semicoherent search is employed: the data are divided into shorter segments, each of which is coherently matched filtered, and the results from each segment combined using a computationally cheaper, less sensitive incoherent method~\cite[e.g.][]{Brady.etal.1998a,Krishnan.etal.2004a,Dergachev.2010a,Pletsch.2010a}.
The most sensitive semicoherent search setup, as a function of e.g.\ the number and length of segments, has been studied in~\cite{Cutler.etal.2005a,Prix.Shaltev.2012a}.

An important component of a wide-parameter-space $\calF$-statistic search for gravitational-wave pulsars is the \emph{metric}, or distance function, associated with the parameter space~\cite{Brady.etal.1998a,Prix.2007a}.
It quantifies how far apart templates may be separated such that any signal in the search parameter space will be recovered with a prescribed maximum \emph{mismatch}, or fractional loss in squared signal-to-noise ratio.
If the metric is not itself a function of the template parameters, the number of templates required to cover the parameter space, and hence the computational cost of the search, can be straightforwardly minimized using the theory of lattices and sphere coverings~\cite{Conway.Sloane.1988,Prix.2007b}.
This is the case for the parameters which describe the gravitational-wave frequency evolution of a pulsar: its frequency at a given reference time, and its frequency time derivatives, or \emph{spindowns}.
It is more difficult, however, to parameterize the pulsar's sky position such that this is true.

Recently, in~\citet[hereafter Paper~I]{Wette.Prix.2013a} we proposed a choice of sky and frequency parameters, with respect to which a close approximation to the metric, the \emph{reduced supersky metric}, is constant.
With respect to previously proposed approximations~\cite{Jaranowski.Krolak.1999a,Astone.etal.2002b,Prix.2007a,Pletsch.Allen.2009a,Pletsch.2010a}, the new metric has no restrictions on the time span of data which can be coherently analyzed, and is numerically well-conditioned which eases its practical use.

Building on the work in~\PaperI, this paper presents further investigations of the reduced supersky metric, with a primary focus on its practical use for template placement.
Section~\ref{sec:background} reviews relevant background information.
In Section~\ref{sec:refin-numer-simul}, we refine numerical simulations, used in~\PaperI\ to test the reduced supersky metric, to more accurately reflect its use in a real search.
In Section~\ref{sec:latt-templ-plac}, we demonstrate that the simulated mismatch predictions of the reduced supersky metric can be realized in a practical search using lattice template placement; this is the first time this has been demonstrated for an all-sky search.
In Section~\ref{sec:number-templates}, we investigate an important property of the reduced supersky metric: the number of templates it predicts are needed to cover the parameter-space of a coherent all-sky search.
Directions for future work are discussed in Section~\ref{sec:discussion}.

\section{Background}\label{sec:background}

This section briefly reviews background information relevant to this paper.
We review the gravitational-wave pulsar signal model and parameter-space metric in Section~\ref{sec:signal-model}, and the supersky and reduced supersky metrics in Section~\ref{sec:supersky-reduc-super}; see~\PaperI\ for further details.
We briefly introduce lattices in Section~\ref{sec:lattices}; see e.g.~\citep{Conway.Sloane.1988} for a comprehensive treatment.

\subsection{The signal model and parameter-space metric}\label{sec:signal-model}

The signal model of a gravitational-wave pulsar~\cite{Jaranowski.etal.1998a}, $h(t, \calA^i, \vec\lambda)$, is a function of four amplitude parameters $\calA^i$, and a number of phase evolution parameters $\vec\lambda$.
The $\calA^i$ are functions of the gravitational-wave strain amplitude $h_0$ and initial phase $\phi_0$, and the pulsar's angles of inclination $\iota$ and polarization $\psi$.
The $\vec\lambda$ are the pulsar's sky position, represented by a unit vector $\vec n$ pointing from the Solar System barycenter (SSB) to the pulsar, frequency $f \equiv \ndot[0] f$ at SSB reference time $t\uref$, and $s\umax$ spindowns $\ndot[s] f \equiv d^s f / d t^s |_{t = t\uref}$.
The signal model can be written as $h(t, \calA, \vec\lambda) = \sum_{i=1}^4 \calA^i h_i(t, \vec\lambda)$, where the $h_i(t, \vec\lambda)$ are four time- and phase-parameter-dependent functions.
We restrict our attention to \emph{isolated} gravitational-wave pulsars, i.e.\ those without a binary companion (but see the discussion in Section~\ref{sec:discussion}).

The $\calF$-statistic~\cite{Jaranowski.etal.1998a,Cutler.Schutz.2005a} matched-filters data from a gravitational-wave detector against the signal model, and further maximizes over the unknown amplitudes $\calA^i$; it is therefore a function of only the phase parameters $\vec\lambda$.
An $\calF$-statistic search computes $\calF(\vec\lambda)$ for a discrete set of values $\{\vec\lambda\Utmpl\}$, which comprise the template bank.
It is unlikely, however, that the parameters $\vec\lambda\Usig$ of any signal in the data will precisely match one of the $\vec\lambda\Utmpl$ in the template bank.
Therefore, the signal will be recovered with a signal-to-noise ratio $\rho(\calA, \vec\lambda\Usig; \vec\lambda\Utmpl)$ lower than for a perfect match $\rho(\calA, \vec\lambda\Usig; \vec\lambda\Usig)$.
The mismatch $\mu\utwoF$ is defined to be~\cite{Prix.2007a,Wette.Prix.2013a}
\begin{equation}
\label{eq:Fstat-mismatch}
\mu\utwoF = \frac{ \rho^2(\calA, \vec\lambda\Usig; \vec\lambda\Usig) - \rho^2(\calA, \vec\lambda\Usig; \vec\lambda\Utmpl) }{ \rho^2(\calA, \vec\lambda\Usig; \vec\lambda\Usig) } \,.
\end{equation}
For small differences $\Delta\vec\lambda = \vec\lambda\Usig - \vec\lambda\Utmpl$, a second-order Taylor expansion of Eq.~\eqref{eq:Fstat-mismatch} yields the metric $\mat g$:
\begin{align}
\label{eq:Fstat-metric}
\mu\utwoF &\approx \Delta\vec\lambda\trsp \frac{-1}{2 \rho^2(\calA, \vec\lambda\Usig; \vec\lambda\Usig)} \left. \frac{\partial \rho^2(\calA, \vec\lambda\Usig; \vec\lambda)}{\partial \vec\lambda} \right|_{\vec\lambda = \vec\lambda\Usig} \Delta\vec\lambda \\
\label{eq:metric-mismatch}
&= \Delta\vec\lambda\trsp \mat g \Delta\vec\lambda \,,
\end{align}
where $\cdot\trsp$ denotes matrix transposition and transformation between row and column vectors.

The elements of $\mat g$ are complicated functions of both amplitude and phase parameters~\cite{Prix.2007a}.
A useful approximation which depends only on the phase parameters is the \emph{phase metric} $\mat g_\phi$~\cite{Brady.etal.1998a,Prix.2007a}, with elements
\begin{equation}
\label{eq:phase-metric-def}
[\mat g_\phi]_{ij} = \big\langle \frac{\partial \phi(t, \vec\lambda)}{\partial \lambda_i} \frac{\partial \phi(t, \vec\lambda)}{\partial \lambda_j} \big\rangle - \big\langle \frac{\partial \phi(t, \vec\lambda)}{\partial \lambda_i} \big\rangle \big\langle \frac{\partial \phi(t, \vec\lambda)}{\partial \lambda_j} \big\rangle \,,
\end{equation}
where $\big\langle x(t) \big\rangle = \int_{t\ustart}^{t\ustart+T} dt \, x(t) / T$, $t\ustart$ is the start time and $T$ is the time span of the data segment being searched.
The function $\phi(t, \vec\lambda)$ denotes the phase of the gravitational-wave pulsar signal in a given detector at time $t$, and is approximately
\begin{equation}
\label{eq:phase-det}
\frac{ \phi(t, \vec\lambda) }{2\pi} \approx \sum_{s=0}^{s\umax} \ndot f \frac{ (t - t\uref)^{s+1} }{ (s+1)! }
+ \frac{ \vec r(t) \cdot \vec n }{c} f\umax \,,
\end{equation}
where $\vec r(t)$ is the detector position relative to the SSB, and $f\umax$ is a constant usually chosen conservatively to be the maximum of the instantaneous frequency $f(t)$ over $T$; see~\PaperI.
If, as is the case for $\ndot f$, the phase $\phi(t, \vec\lambda)$ is linear in a parameter $\lambda_i$, then by Eq.~\eqref{eq:phase-metric-def} $\mat g_\phi$ is independent of $\lambda_i$.

\subsection{The supersky and reduced supersky metrics}\label{sec:supersky-reduc-super}

\PaperI\ proposed adopting the three components of $\vec n$ as sky position parameters; it follows that the phase metric in the coordinates $(\vec n, \ndot f)$, the \emph{supersky} metric $\mat g\uss$, is constant.
A caveat is that this choice of parameters embeds the two-dimensional space of possible sky positions, represented by the two-sphere $|\vec n| = 1$, in the three-dimensional space of possible vectors $\vec n \in \bbR^3$.
Since the physically interesting parameter space is now only a subspace of the full parameter space, template placement is no longer straightforward.

To reduce the dimensionality of $\mat g\uss$, \PaperI\ outlined a procedure which selects a two-dimensional subspace in $\bbR^3$ such that the metric in this subspace, the \emph{reduced supersky} metric $\mat g\urss$, is a close approximation to $\mat g\uss$.
The selected subspace is the plane perpendicular to the eigenvector corresponding to the smallest eigenvalue of $\mat g\ussnn$, the $3{\times}3$ block of $\mat g\uss$ pertaining only to the sky parameters.
Along this axis, the supersky mismatch $\mu\uss$ changes slowest as a function of differences in sky position; dropping this dimension therefore introduces the smallest possible error in an approximation to $\mu\uss$.

In order to reliably perform the above procedure, however, the numerical ill-conditionedness of $\mat g\uss$ must be addressed.
It was found that the ill-conditionedness arises from the near-linear relation, for $T \ll 1$~year, between $\phi(t,\vec\lambda)$ as a function of the orbital motion of the Earth, and as a function of the frequency evolution of the pulsar.
This effect has also been observed and exploited in previous work~\cite{Astone.etal.2002b,Pletsch.Allen.2009a}.
In~\PaperI\ it is used to devise a linear transformation of the coordinates $(\vec n, \ndot f)$ which removes a linear fit to the component of $\mat g\uss$ due to orbital motion by the component due to frequency evolution.

A further linear transformation of the supersky metric removes the correlations between $\vec n$ and $\ndot f$, i.e.\ such that the sky--frequency blocks $\mat g\ussnf$ and $\mat g\ussfn$ of $\mat g\uss$ are zero.
The sky position is then expressed in the eigenbasis of the sky--sky block $\mat g\ussnn$, and the dimension corresponding to the smallest eigenvalue is dropped, yielding the reduced supersky metric $\mat g\urss$.
The metric is constant, and its \emph{condition number}, the ratio of its largest to smallest eigenvalues, is of order unity.
The associated coordinates are $(n\ua, n\ub, \ndot \nu)$, where $n\ua$ and $n\ub$ are the sky coordinates corresponding to the two largest eigenvalues of $\mat g\ussnn$, and the $\ndot \nu$ are frequency and spindown coordinates linear in $\ndot f$ and $\vec n = (n\ua, n\ub, \pm \sqrt{1 - n\ua^2 - n\ub^2})$.

\subsection{Lattices}\label{sec:lattices}

A \emph{lattice} is a set of $n$-dimensional points $\{\vec p\Ulatt\} \subset \bbR^n$ which is closed under vector addition and subtraction, i.e.\ if $\vec p\Ulatt_1$ and $\vec p\Ulatt_2$ are lattice points, then so are $\vec p\Ulatt_1 \pm \vec p\Ulatt_2$.
It follows that a lattice can be generated by a linear transformation from integer vectors $\vec k \in \bbZ^n$ to lattice points $\vec p\Ulatt \in \{\vec p\Ulatt\}$, represented by an $n{\times}n$ \emph{generator matrix} $\mat G$.

The \emph{covering radius} $R$ of a lattice is defined such that every point $\vec p \in \bbR^n$ is within a Euclidean distance $R$ of some lattice point $\vec p\Ulatt \in \{\vec p\Ulatt\}$, i.e.
\begin{equation}
\label{eq:covering-radius-region}
\min_{\vec p\Ulatt \in \{\vec p\Ulatt\}} \|\vec p - \vec p\Ulatt\| \le R^2 \,,
\end{equation}
and no smaller $R$ satisfies this inequality.
The region of $\bbR^n$ defined by Eq.~\eqref{eq:covering-radius-region} is the \emph{covering sphere} centered on the point $\vec p\Ulatt$.

The ratio of the volume contained in a covering sphere to the volume per lattice point, the \emph{normalized thickness} $\theta = R^n / \sqrt{ \det \mat G }$, is a fundamental property of a lattice.
The number of templates needed to cover a given parameter space will be minimized by the lattice with the smallest normalized thickness.

\section{Refined numerical simulations}\label{sec:refin-numer-simul}

In~\PaperI\ the mismatch predictions of the reduced supersky were investigated using numerical simulations.
These simulations generated random parameter offsets $\Delta\vec\lambda$, and examined the difference between the mismatch $\mu\urss = \Delta\vec\lambda\trsp \mat g\urss \Delta\vec\lambda$ predicted by the metric, and the mismatch $\mu_0$ calculated from the $\calF$-statistic using Eq.~\eqref{eq:Fstat-mismatch}; see the Appendix of~\PaperI\ for details.
In Sections~\ref{sec:real-mism-distr}, \ref{sec:fixed-reference-time}, and~\ref{sec:multiple-detectors}, we describe three refinements to the simulations performed in~\PaperI.
The results of the new simulations are presented in Section~\ref{sec:simulation-results}.

\subsection{Realistic mismatch distributions}\label{sec:real-mism-distr}

\begin{figure}
\centering
\subfloat[]{\includegraphics[width=\linewidth]{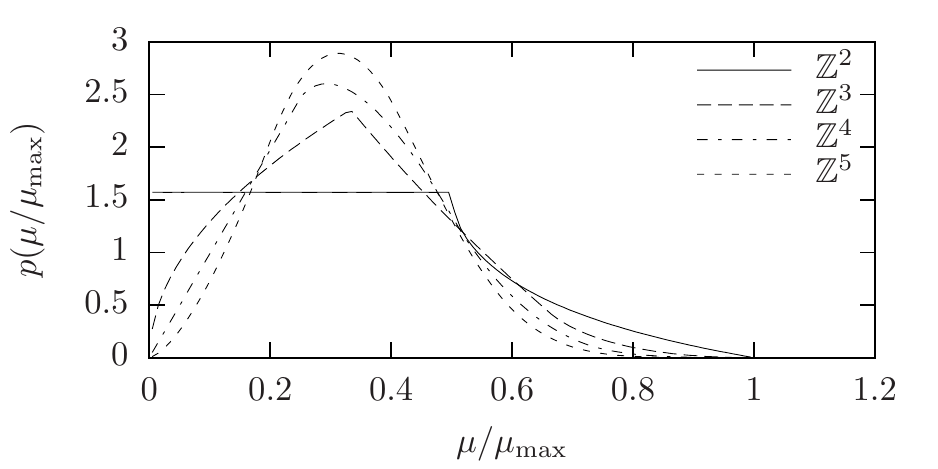}\label{fig:lattice_mism_hgrm_Zn}}\\
\subfloat[]{\includegraphics[width=\linewidth]{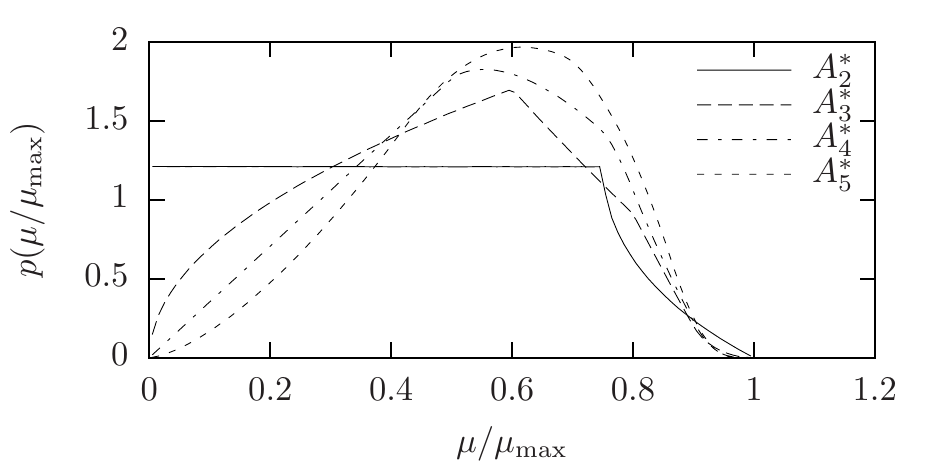}\label{fig:lattice_mism_hgrm_Ans}}
\caption{\label{fig:lattice_mism_hgrm}
Histograms of normalized mismatch $\mu / \mu\umax$ expected from lattice placement using \protect\subref{fig:lattice_mism_hgrm_Zn} $\Zn$ lattices and \protect\subref{fig:lattice_mism_hgrm_Ans} $\Ans$ lattices, with $n = 2$--5.
Each histogram was generated using $10^{9}$ simulated points.
}
\end{figure}

In~\PaperI, the parameter offsets $\Delta\vec\lambda$ were sampled to produce a uniform distribution in $\mu\urss$, up to some maximum $\mu\umax$.
When placing templates using a lattice, however, the expected distribution of mismatches is not uniform, and depends on the geometry of the lattice.
Figure~\ref{fig:lattice_mism_hgrm} plots examples of the mismatch distributions~\cite[cf.][]{Messenger.etal.2009a} expected when placing templates using two lattice families: the $\Zn$ lattices, which generalize the two-dimensional square lattice to higher dimensions; and the $\Ans$ lattices, which generalize the two-dimensional hexagonal lattice.
The $\Ans$ lattices have the smallest known normalized thicknesses in low dimensions~\cite{Conway.Sloane.1988}.

The simulations presented in this paper sample $\Delta\vec\lambda$ to produce the distribution in $\mu\urss$ expected when using an $\Ans$ lattice to place templates in a real search.
This is achieved by choosing a random point $\vec\lambda \in \calP$, then finding the lattice template point $\vec\lambda\Utmpl \in \{\vec\lambda\Utmpl\}$ that minimizes
\begin{equation}
\label{eq:minimize-mismatch}
\mu\urss = (\vec\lambda - \vec\lambda\Utmpl)\trsp \mat g\urss (\vec\lambda - \vec\lambda\Utmpl)
= \Delta\vec\lambda\trsp \mat g\urss \Delta\vec\lambda \,.
\end{equation}
As a consequence, $\mu\urss / \mu\umax$ is distributed according to the mismatch distribution of an $\Ans$ lattice (Fig.~\ref{fig:lattice_mism_hgrm_Ans}).

Let $\mat A$ denote the Cholesky factorization~\citep[e.g.][]{Higham.2002} of $\mat g\urss$, i.e.\ a lower triangular matrix satisfying $\mat g\urss = \mat A \mat A\trsp$.
Then minimizing Eq.~\eqref{eq:minimize-mismatch} is equivalent to minimizing $\| \vec p - \vec p\Ulatt \|$, where $\vec p = \mat A\trsp \vec\lambda$, and $\vec p\Ulatt = \mat A\trsp \vec\lambda\Utmpl$ is the point in the $\Ans$ lattice corresponding to the template $\vec\lambda\Utmpl$.
For many lattices, including $\Ans$, efficient algorithms exist which, given $\vec p$, find the $\vec p\Ulatt$ which minimizes $\| \vec p - \vec p\Ulatt \|$ (see Section~\ref{sec:near-templ-find-idx}).
Once $\vec p\Ulatt$ is found, $\Delta\vec\lambda$ is found via $\Delta\vec\lambda = [\mat A\trsp]\inv (\vec p - \vec p\Ulatt)$, where $\cdot\inv$ denotes matrix inversion.

\subsection{Fixed reference time}\label{sec:fixed-reference-time}

The simulations in~\PaperI\ tested the reduced supersky metric at different start times $t\ustart$ within a 1-year period.
The reference time $t\uref$, which enters the calculation of $\mat g\urss$ via Eq.~\eqref{eq:phase-det}, was always set to $t\uref = t\ustart + 0.5 T$, the mid-time of the data segment for which $\mat g\urss$ was being calculated.

In this paper, we instead fix $t\uref = \mathrm{UTC}$~2007-06-30 00:03:06, and perform the simulations at 25 values of $\Delta t\ustart = t\ustart - t\uref$ from -180 to +180~days in steps of 15~days.
This setup reflects that of a semicoherent search where, in order to combine $\calF$-statistic values from several coherently-analyzed data segments together, it is convenient if the template banks of each data segment (and hence the metrics used to generate them) are defined at the same reference time $t\uref$.
In addition, when $t\uref \ne t\ustart + 0.5 T$ the frequency/spindown off-diagonal elements of the metric, $g\urss(\ndot[s] \nu, \ndot[s\pr] \nu)$ with $s \ne s\pr$, are non-negligibly nonzero; the effect this has on the mismatch predictions of $\mat g\urss$ was not tested in~\PaperI.

\subsection{Multiple detectors}\label{sec:multiple-detectors}

\PaperI\ tested the reduced supersky metric computed at a single detector location, that of the LIGO Hanford detector.
In this paper, we average the reduced supersky metrics computed at the locations of the LIGO Hanford and Livingston detectors.
This is an ad-hoc choice, as the phase metric approximation is defined for a single detector only~\cite{Prix.2007a}.
Nevertheless, the choice proved successful, as demonstrated by the mismatch predictions of the reduced supersky metric presented in the next section.

\subsection{Simulation results}\label{sec:simulation-results}

The simulations presented here follow the procedure outlined in the Appendix of~\PaperI, with the refinements described above.
Simulations are performed at: fixed values of $T$, in steps of 2~days, from 1~to 31~days for first spindown and 11~to 31~days for second spindown; fixed values of $\Delta t\ustart$, as given in Section~\ref{sec:fixed-reference-time}, and fixed values of $f\umax$ from 50~to 1000~Hz, as given in the Appendix of~\PaperI.
Mismatches are compared using their \emph{relative error}, defined following~\PaperI\ to be
\begin{equation}
\label{eq:relerr-def}
\relerr{\mu_a}{\mu_b} = \frac{ \mu_a - \mu_b }{ 0.5( \mu_a + \mu_b ) } \,, \quad \mu_a, \mu_b \ge 0 \,.
\end{equation}

\begin{figure*}
\centering
\subfloat[]{\includegraphics[width=0.49\linewidth]{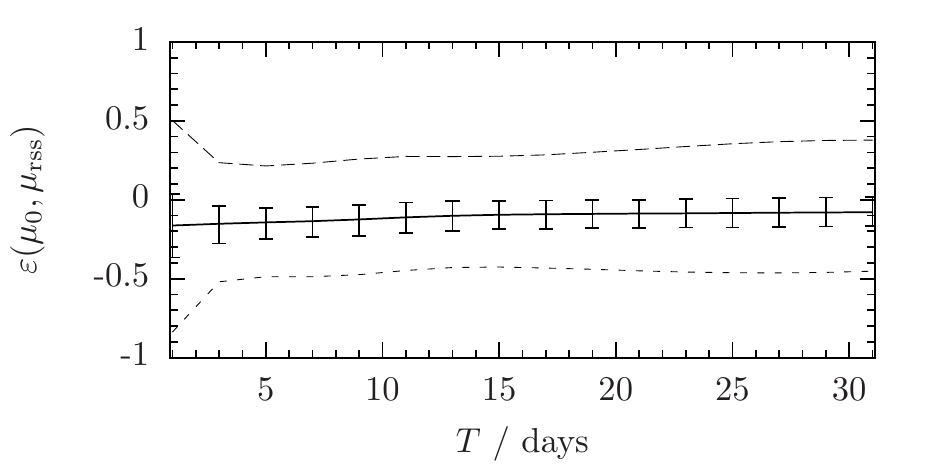}\label{fig:mu_re_T_rss_twoF_f1dot_H1L1_mm0p3_Ans}}
\subfloat[]{\includegraphics[width=0.49\linewidth]{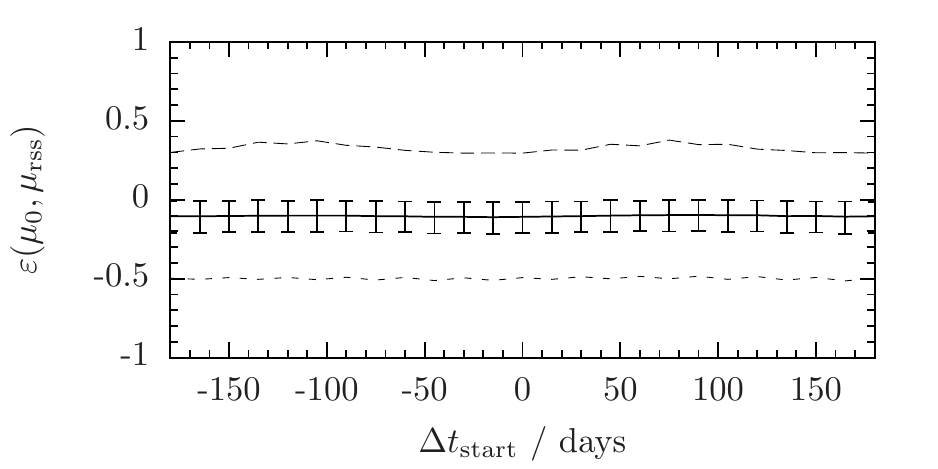}\label{fig:mu_re_dt_rss_twoF_f1dot_H1L1_mm0p3_Ans}}\\
\subfloat[]{\includegraphics[width=0.49\linewidth]{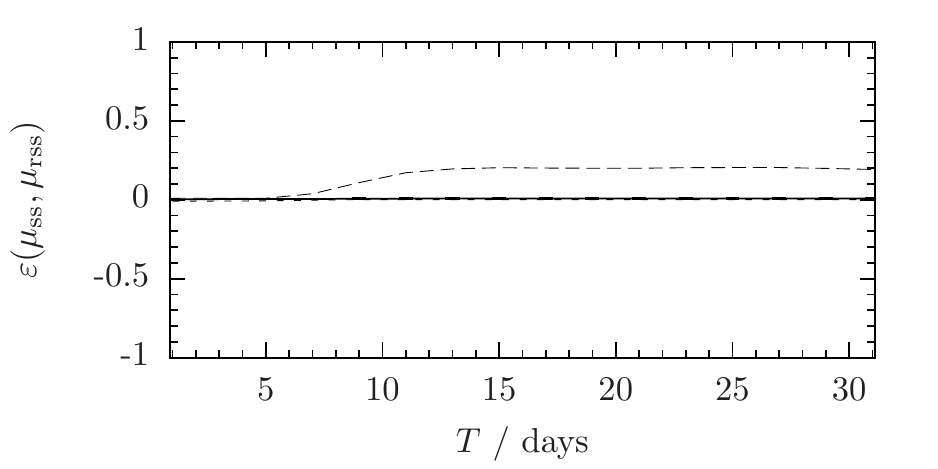}\label{fig:mu_re_T_rss_ss_f1dot_H1L1_mm0p3_Ans}}
\subfloat[]{\includegraphics[width=0.49\linewidth]{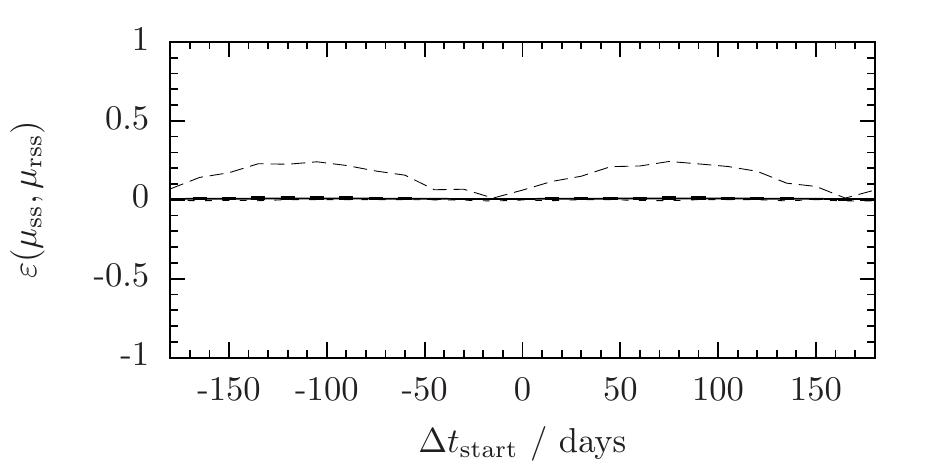}\label{fig:mu_re_dt_rss_ss_f1dot_H1L1_mm0p3_Ans}}
\caption{\label{fig:mu_re_f1dot}
Relative errors between $\mu\urss$ and $\mu\utwoF$ (top row), and between $\mu\urss$ and $\mu\uss$ (bottom row).
Only first spindown is used.
Plotted are the median (solid line), the 25th--75th percentile range (error bars), and the 2.5th (short-dashed line) and 97.5th (long-dashed line) percentiles of relative errors: as a function of $T$, averaged over $\Delta t\ustart$ and $f\umax$ (left column); and as a function of $\Delta t\ustart$, averaged over $T$ and $f\umax$ (right column).
An $\Ans$ lattice is used to generate random parameter offsets with $\mu\urss \le 0.3$.
}
\end{figure*}

\begin{figure*}
\centering
\subfloat[]{\includegraphics[width=0.49\linewidth]{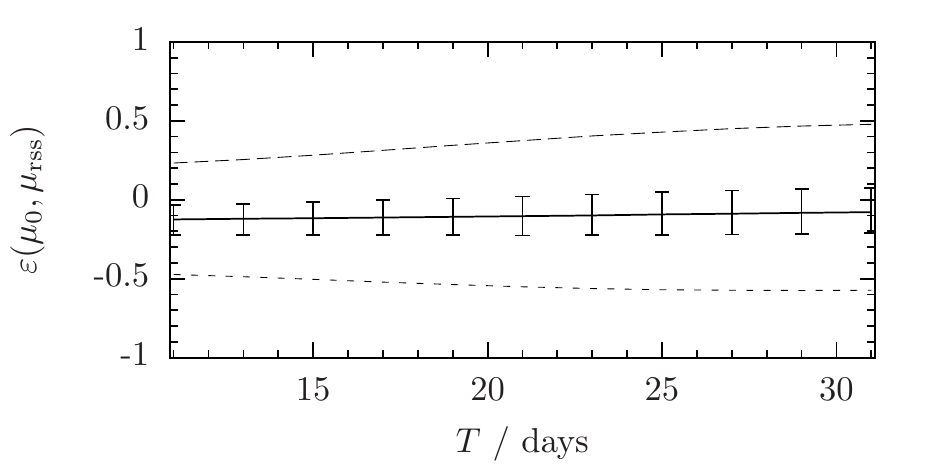}\label{fig:mu_re_T_rss_twoF_f2dot_H1L1_mm0p3_Ans}}
\subfloat[]{\includegraphics[width=0.49\linewidth]{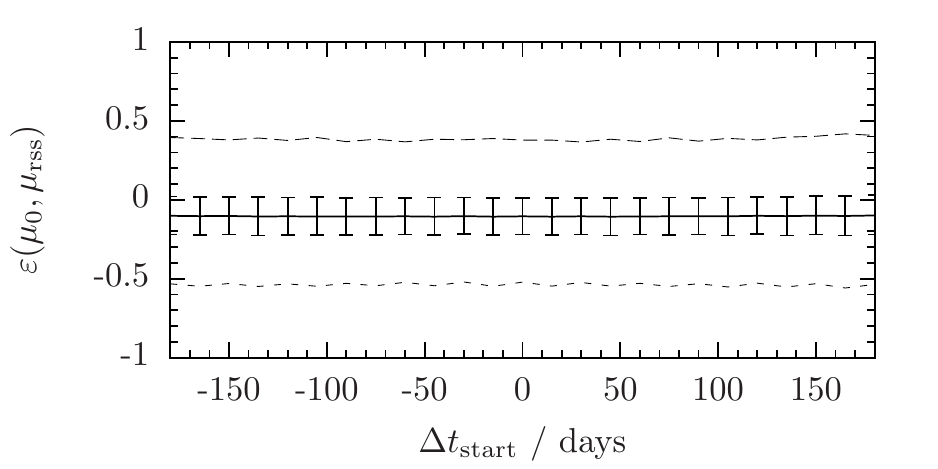}\label{fig:mu_re_dt_rss_twoF_f2dot_H1L1_mm0p3_Ans}}\\
\subfloat[]{\includegraphics[width=0.49\linewidth]{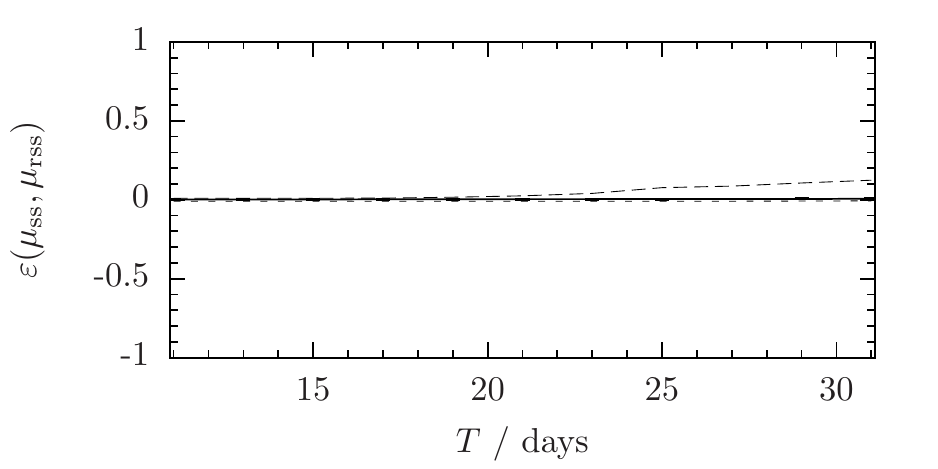}\label{fig:mu_re_T_rss_ss_f2dot_H1L1_mm0p3_Ans}}
\subfloat[]{\includegraphics[width=0.49\linewidth]{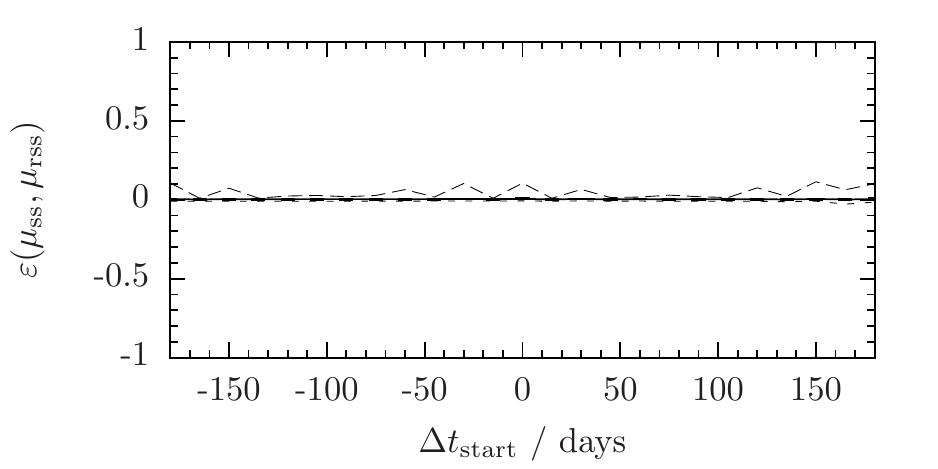}\label{fig:mu_re_dt_rss_ss_f2dot_H1L1_mm0p3_Ans}}
\caption{\label{fig:mu_re_f2dot}
Relative errors between $\mu\urss$ and $\mu\utwoF$ (top row), and between $\mu\urss$ and $\mu\uss$ (bottom row).
Both first and second spindown are used.
Plotted are the median (solid line), the 25th--75th percentile range (error bars), and the 2.5th (short-dashed line) and 97.5th (long-dashed line) percentiles of relative errors: as a function of $T$, averaged over $\Delta t\ustart$ and $f\umax$ (left column); and as a function of $\Delta t\ustart$, averaged over $T$ and $f\umax$ (right column).
An $\Ans$ lattice is used to generate random parameter offsets with $\mu\urss \le 0.3$.
}
\end{figure*}

Figures~\ref{fig:mu_re_f1dot} and~\ref{fig:mu_re_f2dot} plot the relative errors $\relerr{\mu\utwoF}{\mu\urss}$ and $\relerr{\mu\uss}{\mu\urss}$ between mismatches $\mu\urss$ and $\mu\uss$ predicted by the reduced supersky and supersky metrics, $\mat g\urss$ and $\mat g\uss$ respectively, and the mismatch $\mu\utwoF$ calculated from the $\calF$-statistic.
The simulations presented in Fig.~\ref{fig:mu_re_f1dot} test metrics computed at time spans $T$ from 1~to 31~days using random offsets in the sky coordinates $(n\ua, n\ub)$, frequency $\nu$, and first spindown $\dot\nu$; those presented in Fig.~\ref{fig:mu_re_f2dot} use $T$ from 11~to 31~days and include additional random offsets in second spindown $\ddot\nu$.
Figures~\ref{fig:mu_re_T_rss_twoF_f1dot_H1L1_mm0p3_Ans} and~\ref{fig:mu_re_T_rss_ss_f1dot_H1L1_mm0p3_Ans} are comparable to Figs.~2f and 14 of~\PaperI, while Figs.~\ref{fig:mu_re_T_rss_twoF_f2dot_H1L1_mm0p3_Ans} and~\ref{fig:mu_re_T_rss_ss_f2dot_H1L1_mm0p3_Ans} are comparable to Figs.~17a and~17b of that paper.

The simulations presented here show comparable, even improved, mismatch predictions by the reduced supersky metric compared to~\PaperI, despite using a slightly larger maximum mismatch of 0.3 (cf.\ 0.2 in~\PaperI).
For example, Fig.~2f of~\PaperI\ shows $\relerr{\mu\utwoF}{\mu\urss} \sim -0.5$ at $T = 1$~day, whereas Fig.~\ref{fig:mu_re_T_rss_twoF_f1dot_H1L1_mm0p3_Ans} shows $\relerr{\mu\utwoF}{\mu\urss} \sim -0.15$ at the same $T$.
This is expected; as seen in Fig.~\ref{fig:lattice_mism_hgrm}, the mismatch distribution expected from using a $\Ans[4]$ lattice includes fewer large mismatches (e.g.\ $\mu/\mu\umax \gtrsim 0.8$) than would a uniform distribution.
We expect the metric to perform worse at larger mismatches, due to the deterioration of the metric approximation to the $\calF$-statistic mismatch (see Sec.~IV~A and Fig.~7 of~\PaperI).
The small relative errors seen in the bottom rows of Figs.~\ref{fig:mu_re_f1dot} and~\ref{fig:mu_re_f2dot} confirm the close agreement between the supersky $\mat g\uss$ and reduced supersky $\mat g\urss$ metrics seen in Fig.~14 of~\PaperI.

No deterioration in the mismatch predictions of the reduced supersky metric as a function of $\Delta t\ustart$ is observed in Figs.~\ref{fig:mu_re_dt_rss_twoF_f1dot_H1L1_mm0p3_Ans}, \ref{fig:mu_re_dt_rss_ss_f1dot_H1L1_mm0p3_Ans}, \ref{fig:mu_re_dt_rss_twoF_f2dot_H1L1_mm0p3_Ans}, and~\ref{fig:mu_re_dt_rss_ss_f2dot_H1L1_mm0p3_Ans}.
This indicates that the derivation of the reduced supersky metric is robust to the difference between reference time $t\uref$ and segment start time $t\ustart$.

\begin{figure}
\centering
\subfloat[]{\includegraphics[width=\linewidth]{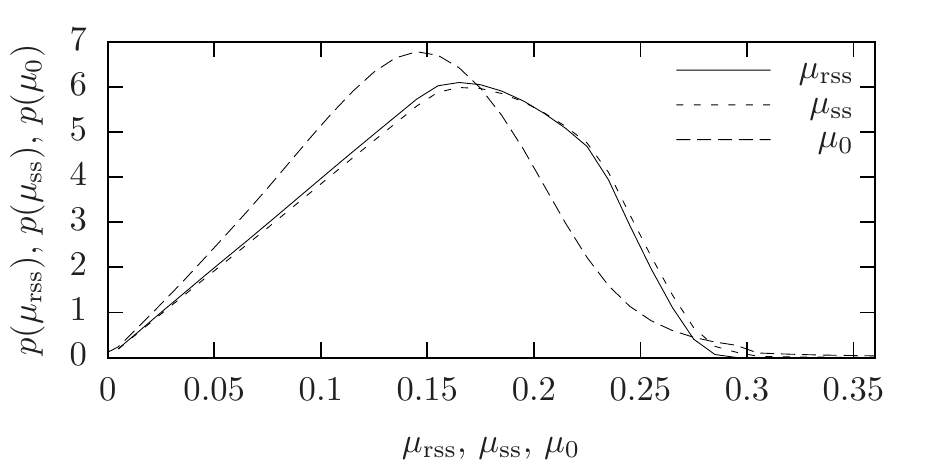}\label{fig:mu_hgrm_f1dot_H1L1_mm0p3_Ans}}\\
\subfloat[]{\includegraphics[width=\linewidth]{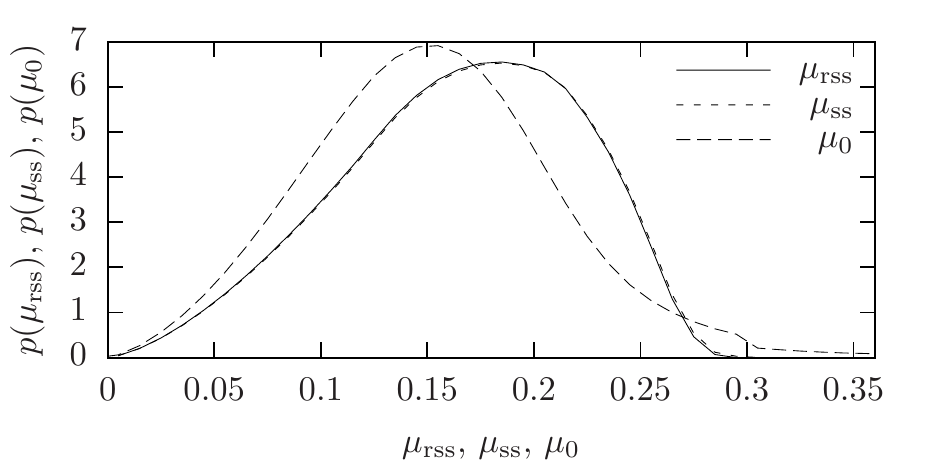}\label{fig:mu_hgrm_f2dot_H1L1_mm0p3_Ans}}
\caption{\label{fig:mu_hgrm}
Histograms of mismatches in $\mu\urss$, $\mu\uss$, and $\mu\utwoF$, averaged over all simulated values of $T$, $\Delta t\ustart$, and $f\umax$.
Only first spindown is used in \protect\subref{fig:mu_hgrm_f1dot_H1L1_mm0p3_Ans};
both first and second spindown are used in \protect\subref{fig:mu_hgrm_f2dot_H1L1_mm0p3_Ans}.
An $\Ans$ lattice is used to generate random parameter offsets with $\mu\urss \le 0.3$.
}
\end{figure}

Figure~\ref{fig:mu_hgrm} compares the distributions of mismatch in $\mu\urss$, $\mu\utwoF$, and $\mu\uss$, averaged over all simulation parameters.
The mismatch distributions of $\mu\urss$ closely resemble the desired mismatch distributions for an $\Ans$ lattice; see Fig.~\ref{fig:lattice_mism_hgrm_Ans} with $n = 4$ (first spindown only) and $n = 5$ (both first and second spindown).
Given the close agreement between the reduced supersky and supersky metrics seen in Figs.~\ref{fig:mu_re_T_rss_ss_f1dot_H1L1_mm0p3_Ans}, \ref{fig:mu_re_dt_rss_ss_f1dot_H1L1_mm0p3_Ans}, \ref{fig:mu_re_T_rss_ss_f2dot_H1L1_mm0p3_Ans}, and~\ref{fig:mu_re_dt_rss_ss_f2dot_H1L1_mm0p3_Ans}, it is expected that the mismatch distributions of $\mu\uss$ resemble those of $\mu\urss$.

Likewise, the differences between the $\mu\utwoF$ and $\mu\urss$ mismatch distributions seen in Fig.~\ref{fig:mu_hgrm} are also expected, given that the reduced supersky metric does not perfectly predict the $\calF$-statistic mismatch (Figs.~\ref{fig:mu_re_T_rss_twoF_f1dot_H1L1_mm0p3_Ans}, \ref{fig:mu_re_dt_rss_twoF_f1dot_H1L1_mm0p3_Ans}, \ref{fig:mu_re_T_rss_twoF_f2dot_H1L1_mm0p3_Ans}, and~\ref{fig:mu_re_dt_rss_twoF_f2dot_H1L1_mm0p3_Ans}).
That the $\mu\utwoF$ distributions peak at lower mismatches than those of $\mu\urss$ is consistent with the reduced supersky metric overestimating the $\calF$-statistic mismatch; see the discussion of Fig.~7 in~\PaperI.
On the other hand, the means of the distributions are very similar: for first spindown only, the mean $\mu\urss$ mismatch is 0.15 and the mean $\mu\utwoF$ mismatch is 0.14; for both first and second spindown, the means are 0.17 and 0.15 respectively.

\section{Lattice template placement}\label{sec:latt-templ-plac}

In the previous section, we confirmed that the $\calF$-statistic mismatch is well-predicted by the reduced supersky metric, when the distribution of $\mu\urss$ resembles that expected for lattice template placement.
In this section, we demonstrate that such mismatch distributions are realized by a practical implementation~\footnote{
The implementation is available as the \textsc{LatticeTiling} module of the \textsc{LALPulsar} library, a part of the \textsc{LALSuite} software for gravitational-wave data analysis; see \url{https://www.lsc-group.phys.uwm.edu/daswg/projects/lalsuite.html}.
} of a lattice template bank.
The implementation is based on one developed in~\cite{Wette.2009a}, which was used in the gravitational-wave pulsar search presented in~\cite{Abadie.etal.2010b}; lattice template placement is also discussed in~\cite{Prix.2007b}.
Constituent parts of the implementation are described in Sections~\ref{sec:param-space-repr}, \ref{sec:latt-templ-gener}, and~\ref{sec:near-templ-find-idx}; tests of the implementation are presented in Section~\ref{sec:latt-templ-plac-test}.
An issue pertaining to the coverage of parameter-space boundaries is discussed in Section~\ref{sec:stairc-bound-templ}.

\subsection{Parameter-space representation}\label{sec:param-space-repr}

\begin{figure}
\centering
\includegraphics[width=\linewidth]{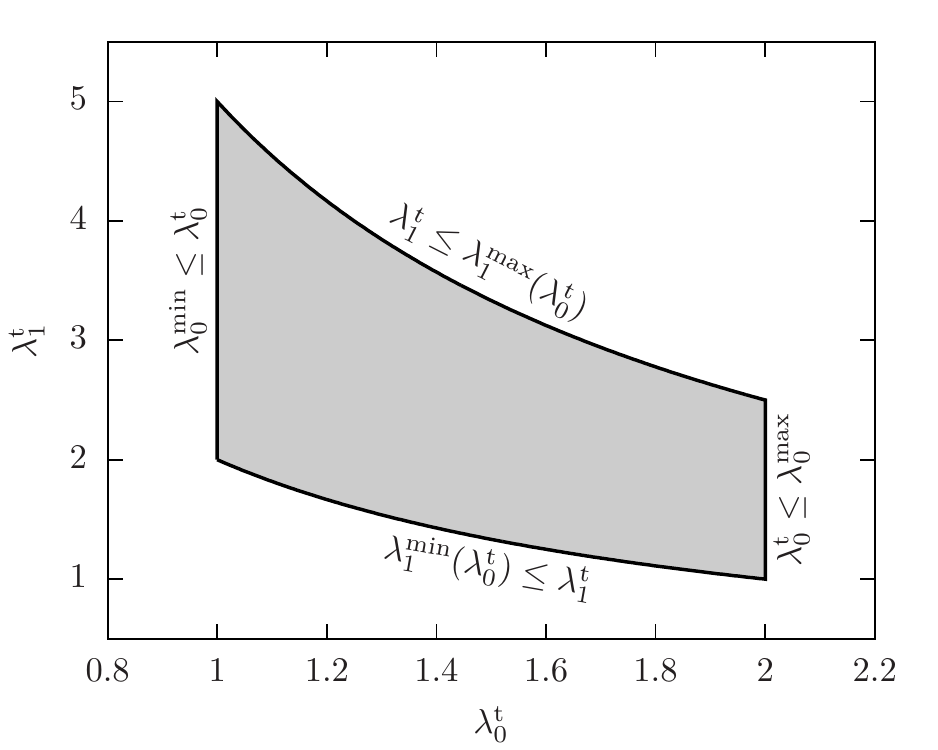}
\caption{\label{fig:parameter_space_example}
Illustration of a two-dimensional parameter space $\calP$ (gray shaded area) whose boundaries (solid lines) are represented, per Eq.~\eqref{eq:param-space-lambda}, by $1 \le \lambda\Utmpl_0 \le 2$ and $2 / \lambda\Utmpl_0 \le \lambda\Utmpl_1 \le 5 / \lambda\Utmpl_0$.
Each boundary is labeled by its defining inequality.
}
\end{figure}

We consider an $n$-dimensional parameter space $\calP \subset \bbR^n$, within which templates are points $\vec\lambda\Utmpl = (\lambda\Utmpl_0, \lambda\Utmpl_1, \dots, \lambda\Utmpl_{n-1}) \in \bbR^n$.
We prescribe that $\calP$ be represented by inequalities of the form:
\begin{equation}
\label{eq:param-space-lambda}
\begin{gathered}
\lambda\Umin_0 \le \lambda\Utmpl_0 \le \lambda\Umax_0 \,, \\
\lambda\Umin_1(\lambda\Utmpl_0) \le \lambda\Utmpl_1 \le \lambda\Umax_1(\lambda\Utmpl_0) \,, \\
\lambda\Umin_2(\lambda\Utmpl_0, \lambda\Utmpl_1) \le \lambda\Utmpl_2 \le \lambda\Umax_2(\lambda\Utmpl_0, \lambda\Utmpl_1) \,, \\
\dots, \\
\lambda\Umin_{n-1}(\dots, \lambda\Utmpl_{n-2}) \le \lambda\Utmpl_{n-1} \le \lambda\Umax_{n-1}(\dots, \lambda\Utmpl_{n-2}) \,.
\end{gathered}
\end{equation}
The lowest dimension $\lambda\Utmpl_0$ is bounded by two constants $\lambda\Umin_0$ and $\lambda\Umax_0$; the dimension $\lambda\Utmpl_1$ by two functions $\lambda\Umin_1(\lambda\Utmpl_0)$ and $\lambda\Umax_1(\lambda\Utmpl_0)$, depending only on $\lambda\Utmpl_0$; the dimension $\lambda\Utmpl_2$ by two functions $\lambda\Umin_2(\lambda\Utmpl_0, \lambda\Utmpl_1)$ and $\lambda\Umax_2(\lambda\Utmpl_0, \lambda\Utmpl_1)$, depending only on $\lambda\Utmpl_0$ and $\lambda\Utmpl_1$; and so on up to $\lambda\Utmpl_{n-1}$.
The motivation for this representation is presented in the next section.
Figure~\ref{fig:parameter_space_example} shows an example of a parameter space which is described in this form.

The region of $\calP$ covered by a template $\vec\lambda\Utmpl$ is the set of points $\{\vec\lambda\}$ such that
\begin{equation}
\label{eq:metric-ellipse-region}
(\vec\lambda - \vec\lambda\Utmpl)\trsp \mat g\urss (\vec\lambda - \vec\lambda\Utmpl) \le \mu\umax \,;
\end{equation}
this expression describes an $n$-dimensional \emph{metric ellipse}.
The \emph{metric ellipse bounding box} of $\mat g\urss$ is the smallest $n$-dimensional coordinate box which contains the metric ellipse; its widths $\beta_i$ in each dimension $i = 0, \dots, n-1$ are the minimum required to satisfy $|\lambda_i - \lambda\Utmpl_i| \le 0.5 \beta_i$, for all points $\vec\lambda$ satisfying Eq.~\eqref{eq:metric-ellipse-region}.
The widths $\vec\beta$ are computed from the metric via \citep[e.g.][]{Wette.2009a}
\begin{equation}
\label{eq:bounding-box}
\beta_i = 2 \sqrt{ \mu\umax [\mat g\urss\inv]_{ii} } \,,
\end{equation}
where $[\mat g\urss\inv]_{ii}$ is the $i$th diagonal element of $\mat g\urss\inv$.

\begin{figure}
\centering
\includegraphics[width=\linewidth]{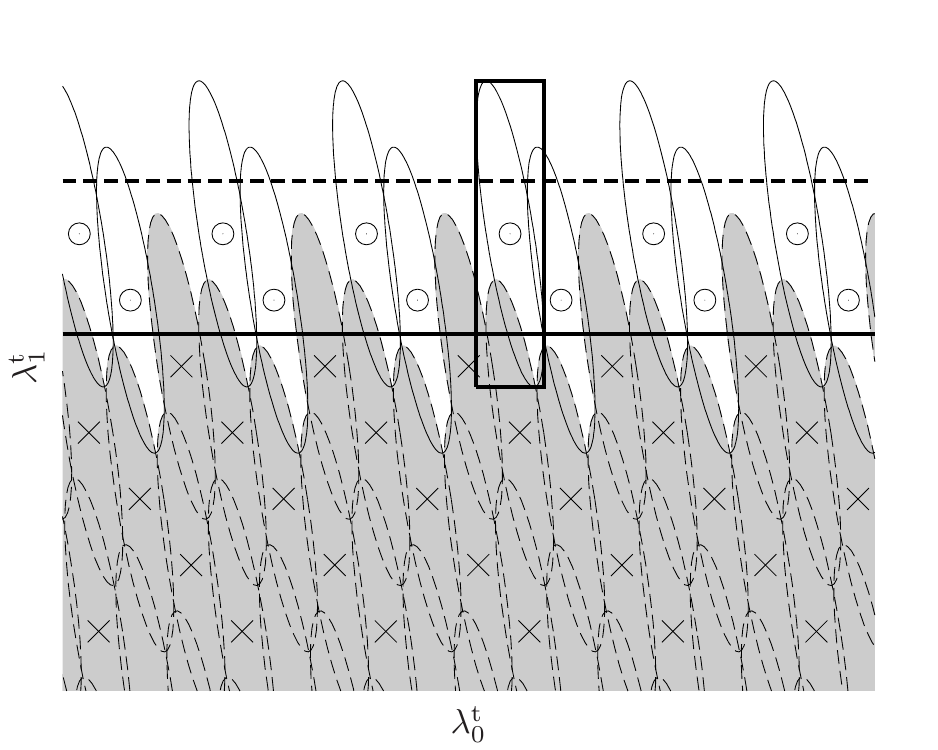}
\caption{\label{fig:boundary_covering_diagram}
Illustration of the boundary of a two-dimensional parameter space $\calP$.
Templates (crosses, dashed ellipses) are laid within the bound $\lambda\Utmpl_1 \le \lambda\Umax_1$ (solid line); the gray shaded area indicates where $\calP$ is covered by these templates.
The white areas within the $\lambda\Utmpl_1 \le \lambda\Umax_1$ bound are covered by laying templates outside the parameter space (circles, solid ellipses), up to $\lambda\Utmpl_1 \le \lambda\Umax_1 + 0.5\beta_1$ (dashed line), where $\beta_1$ is the height of the metric ellipse bounding box (solid box).
}
\end{figure}

Care must be taken at the boundaries of $\calP$ to ensure that it is completely covered.
Figure~\ref{fig:boundary_covering_diagram} illustrates a common situation where simply laying templates inside the given bounds of $\calP$ does not fully cover it.
By extending the boundaries of $\calP$ by half of the extent of the bounding box, given by Eq.~\eqref{eq:bounding-box}, complete coverage is achieved.

\subsection{Lattice template generation}\label{sec:latt-templ-gener}

The template points $\vec\lambda\Utmpl$ are generated from integer vectors $\vec k \in \bbZ^n$ by
\begin{equation}
\label{eq:template-gen}
\vec\lambda\Utmpl = \mat T \vec k \,,
\end{equation}
where the matrix $\mat T$ transforms from $\bbZ^n$ to $\calP$.
Each template point $\vec\lambda\Utmpl$ is placed at a vertex of the lattice.
It can be shown~\cite{Wette.2009a} that, if $\mat T$ is a lower triangular matrix, the bounds on $\vec k = (k_0, k_1, \dots. k_{n-1})$ can be represented in the same manner as the bounds on $\vec\lambda\Utmpl$ in Eq.~\eqref{eq:param-space-lambda}:
\begin{equation}
\label{eq:param-space-k}
\begin{gathered}
k_0\Umin \le k_0 \le k_0\Umax \,, \\
k_1\Umin(k_0) \le k_1 \le k_1\Umax(k_0) \,, \\
k_2\Umin(k_0, k_1) \le k_2 \le k_2\Umax(k_0, k_1) \,, \\
\dots, \\
k_{n-1}\Umin(\dots, k_{n-2}) \le k_{n-1} \le k_{n-1}\Umax(\dots, k_{n-2}) \,.
\end{gathered}
\end{equation}
This representation suggests an algorithm for iterative generation of templates using $n$ nested loops.
The outer-most loop generates integer values of $k_0$ between $k_0\Umin$ and $k_0\Umax$; for each $k_0$, the next inner loop calculates the bounds $k_1\Umin(k_0)$ and $k_1\Umax(k_0)$ and generates integer values of $k_1$ between the two bounds; and so on until the inner-most loop which, for each $k_0, \dots, k_{n-2}$, generates integer values of $k_{n-1}$ between $k_{n-1}\Umin(k_0, \dots, k_{n-2})$ and $k_{n-1}\Umax(k_0, \dots, k_{n-2})$.
This algorithm ensures that all of $\calP$ is visited, regardless of its geometry, and is the motivation for the representation of $\calP$ given by Eq.~\eqref{eq:param-space-lambda}.

We now derive the lower triangular matrix $\mat T$ which transforms integers $\vec k$ to template points $\vec\lambda\Utmpl$.
A $n$-dimensional lattice embedded in $m$-dimensional space is generated by a $m{\times}n$ matrix~\footnote{
Some lattices, e.g.\ $\Ans$, are conveniently represented in $m$-dimensional space, by an $m{\times}m$ matrix $\mat G\pr$, of which the lattice occupies an $n$-dimensional subspace.
Hence only $n$ columns of $\mat G\pr$ are linearly independent, and we take $\mat G$ to be the $m{\times}n$ matrix comprising linearly independent columns of $\mat G\pr$.
} $\mat G$, $m \ge n$, such that
\begin{equation}
\label{eq:lattice-gen-G}
\vec p\Ulatt[\pr] = \mat G \vec k \,,
\end{equation}
where $\vec k \in \bbZ^n$ are integer vectors, and $\vec p\Ulatt[\pr] \in \bbR^m$ are the lattice points embedded in $m$-dimensional space.
To find a representation of the lattice in $n$-dimensional space, we compute the QL factorization~\footnote{
The QL factorization may be calculated using the more commonly implemented QR factorization~\citep[e.g.][]{Higham.2002} via $E(\mat G) = \mat Q\pr \mat R$, $E(\mat Q\pr) = \mat Q$, $E(\mat R) = \mat L\pr$, where $E(\cdot)$ reverses the order of both the rows and columns of its argument.
} of $\mat G$:
\begin{equation}
\label{eq:lattice-gen-QL}
\mat G = \mat Q \mat L\pr \,,
\end{equation}
where $\mat Q$ is an $m{\times}m$ orthogonal matrix, and $\mat L\pr$ is an $m{\times}n$ matrix with zeros above the $(m - n)$th subdiagonal.

The matrix $\mat Q$ denotes an overall rotation, which does not affect the lattice's covering properties, and can therefore be discarded.
The top $(m - n)$ rows of $\mat L\pr$ are zero, and are therefore dropped; let $\mat L$ denote the lower triangular matrix comprising the remaining, lower $n$ rows of $\mat L\pr$.
The lattice can then be generated using
\begin{equation}
\label{eq:lattice-gen-Lpr}
\vec p\Ulatt = \mat L \vec k \,,
\end{equation}
where $\vec p\Ulatt \in \bbR^n$ are the lattice points, now embedded in $n$-dimensional space.

Finally, we require a transformation from lattice points $\{\vec p\Ulatt\}$ to templates $\{\vec\lambda\Utmpl\}$ such that Eq.~\eqref{eq:metric-ellipse-region} is satisfied, i.e.\ each point $\vec\lambda \in \calP$ is within a mismatch $\mu\umax$ of some template $\vec\lambda\Utmpl$.
Let
\begin{equation}
\label{eq:template-gen-p-to-lambda}
\vec\lambda = \frac{ \sqrt{\mu\umax} }{ R } \, \mat B \vec p \,, \quad
\vec\lambda\Utmpl = \frac{ \sqrt{\mu\umax} }{ R } \, \mat B \vec p\Ulatt \,,
\end{equation}
where $R$ is the covering radius of the lattice, and $\mat B$ is the Cholesky factorization of $\mat g\urss\inv$, i.e.\ a lower triangular matrix satisfying $\mat g\urss\inv = \mat B \mat B\trsp$.
Substituting this and Eq.~\eqref{eq:template-gen-p-to-lambda} into Eq.~\eqref{eq:metric-ellipse-region} gives
\begin{equation}
\frac{ \mu\umax }{ R^2 } \big[ \mat B ( \vec p - \vec p\Ulatt ) \big]\trsp \big[ \mat B \mat B\trsp ]\inv \big[ \mat B ( \vec p - \vec p\Ulatt ) \big] \le \mu\umax \,,
\end{equation}
which simplifies to Eq.~\eqref{eq:covering-radius-region}, the definition of a covering sphere (see Section~\ref{sec:lattices}).
Since Eq.~\eqref{eq:covering-radius-region} is always satisfied by the lattice points $\{\vec p\Ulatt\}$, Eq.~\eqref{eq:metric-ellipse-region} will also always be satisfied by the templates $\{\vec\lambda\Utmpl\}$.
The required lower triangular matrix $\mat T$ is therefore
\begin{equation}
\label{eq:template-gen-T}
\mat T = \frac{ \sqrt{\mu\umax} }{ R } \, \mat B \mat L \,.
\end{equation}

\begin{figure*}
\includegraphics[width=\linewidth]{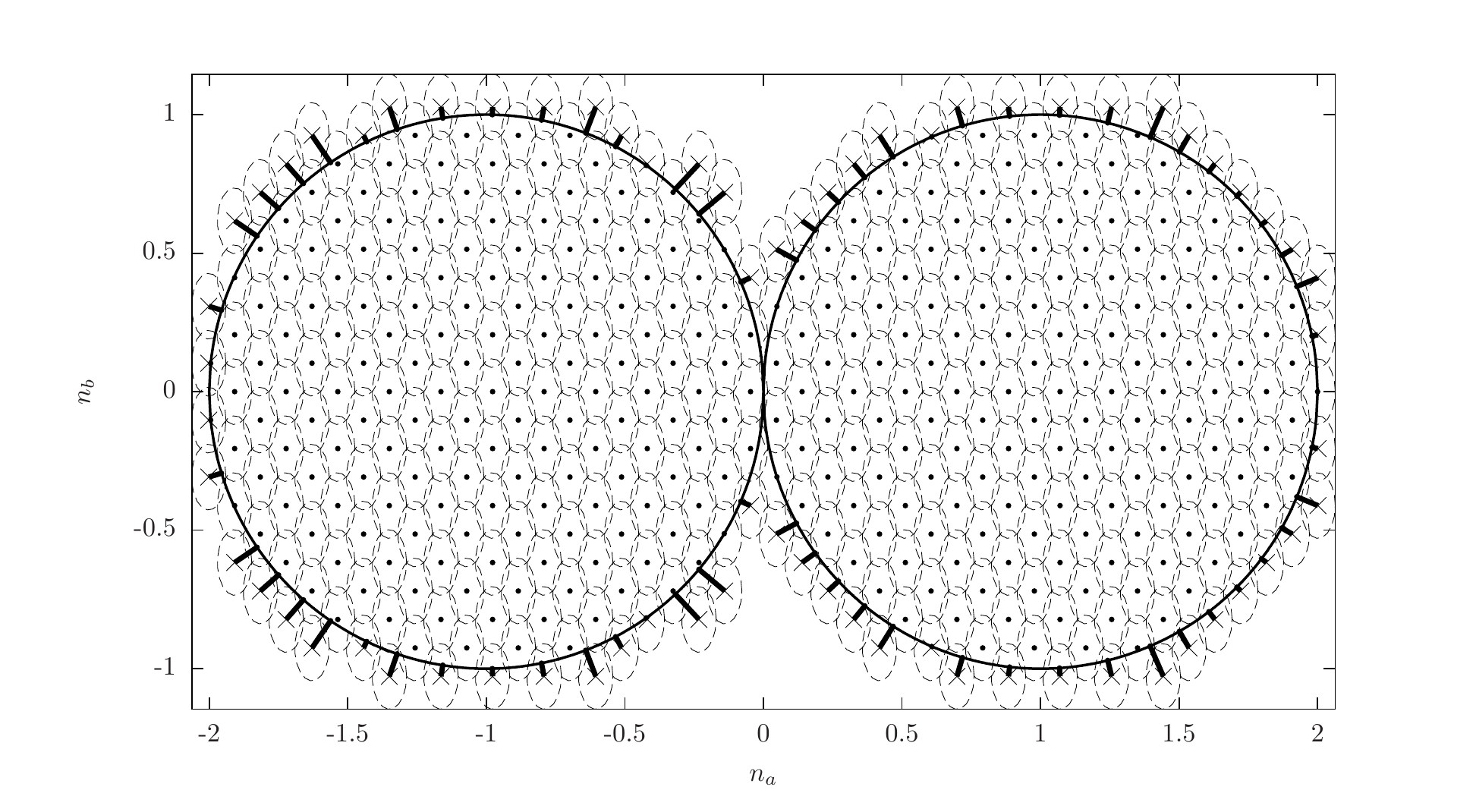}
\caption{\label{fig:tmpl_example_sky_bank}
Example of a lattice template bank in reduced supersky coordinates $(n\ua, n\ub)$, with $T = 1$~day, $\Delta t\ustart = 0$, and $\nu\umax = 100$~Hz.
Templates and their metric ellipses are plotted as points and dashed lines respectively.
Templates, plotted as crosses, which lie outside the parameter-space boundaries (circles) are moved radially onto the nearest boundary (thick lines) when converted to physical coordinates $(\alpha, \delta)$.
An $\Ans$ lattice is used to generate the template bank with $\mu\urss \le 0.3$.
}
\end{figure*}

\begin{figure}
\includegraphics[width=\linewidth]{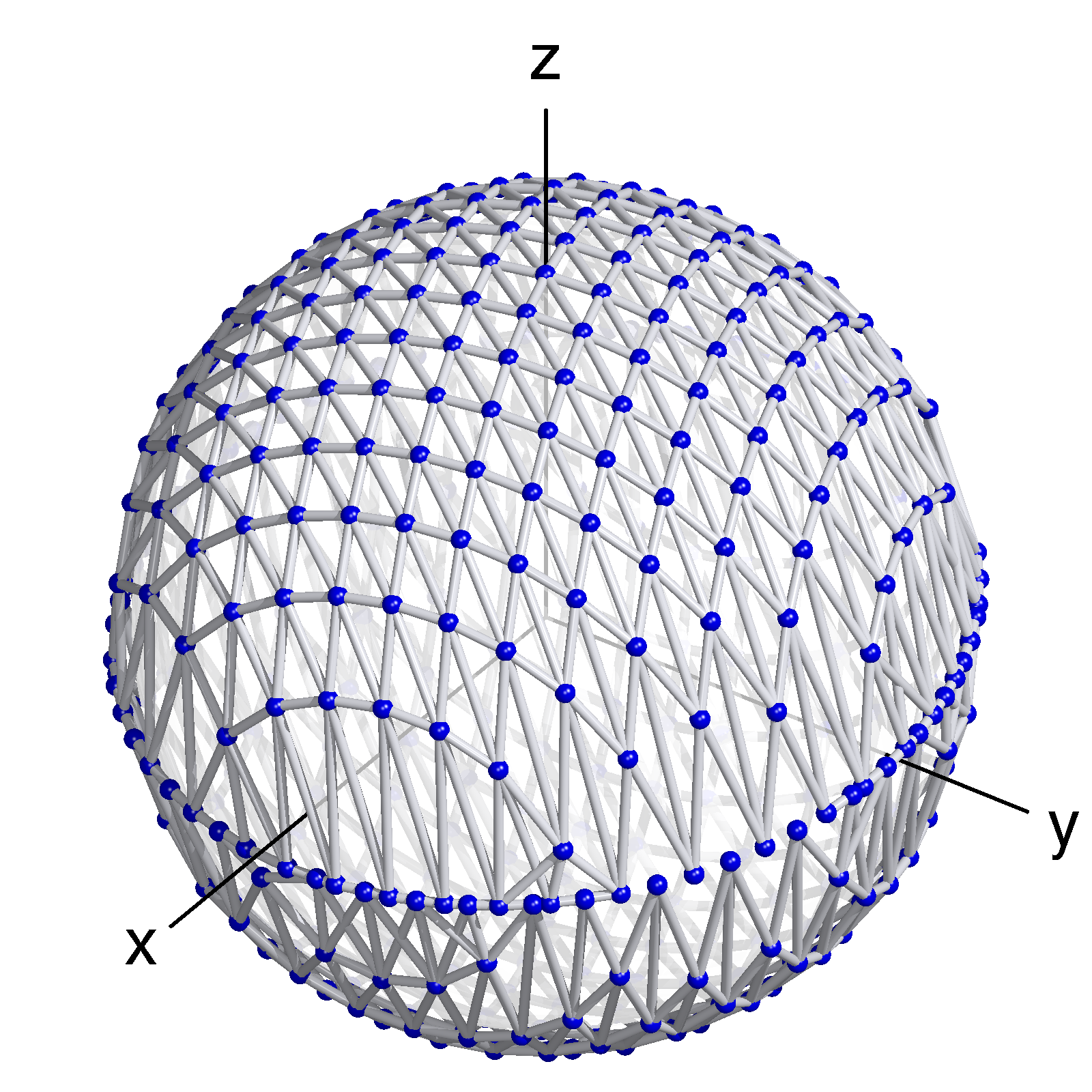}
\caption{\label{fig:tmpl_example_sky_sphere}
The projection of the example lattice template bank from Fig.~\ref{fig:tmpl_example_sky_bank} onto the sky sphere $|\vec n| = 1$.
Templates are plotted as points; lines are drawn from each template to its (up to) six nearest neighbors in the $\Ans$ lattice.
}
\end{figure}

Figure~\ref{fig:tmpl_example_sky_bank} plots an example lattice template bank in the reduced supersky metric sky coordinates $(n\ua, n\ub)$.
The parameter space comprises two unit disks centered on $n\ua = \pm 1, n\ub = 0$, one for each hemisphere of the sky.
To convert from reduced supersky to physical sky coordinates, e.g.\ right ascension $\alpha$ and declination $\delta$, sky positions are converted to supersky coordinates $\vec n = (n\ux, n\uy, n\uz)$, and projected onto the sky sphere $|\vec n| = 1$, as illustrated in Fig.~\ref{fig:tmpl_example_sky_sphere}.
Boundary templates which lie outside the reduced supersky parameter space are projected radially onto the parameter-space boundary, as shown in Fig.~\ref{fig:tmpl_example_sky_bank}.
Finally, $(\alpha, \delta)$ are calculated from $\vec n = (\cos\alpha \cos\delta, \sin\alpha \cos\delta, \sin\delta)$.

\subsection{Nearest template finding and indexing}\label{sec:near-templ-find-idx}

Given a lattice template bank $\{\vec\lambda\Utmpl\}$, we would like to be able to find the nearest template $\vec\lambda\Utmpl(\vec\lambda) \in \{\vec\lambda\Utmpl\}$ to any given point $\vec\lambda \in \calP$, from which we can calculate $\vec\Delta\lambda = \vec\lambda - \vec\lambda\Utmpl(\vec\lambda)$ and the mismatch $\mu\urss$ via Eq.~\eqref{eq:metric-mismatch}.
As discussed in Section~\ref{sec:discussion}, this facility would also be needed by a semicoherent search.
Efficient algorithms for finding $\vec\lambda\Utmpl(\vec\lambda)$ are specialized to the type of lattice being used.
For example, the most efficient algorithm for $\Zn$ lattices is
\begin{equation}
\label{eq:nearest-point-Zn}
\vec\lambda\Utmpl(\vec\lambda)|_{\Zn} = \mat T \left\lfloor \mat T\inv \vec\lambda \right\rceil \,,
\end{equation}
where the operation $\lfloor\cdot\rceil$ rounds each vector element to the nearest integer.
For $\Ans$ lattices, several efficient algorithms have been proposed; see~\cite{Conway.Sloane.1988,McKilliam.etal.2008a} and references therein.

We implement the algorithm described~\footnote{
We note that line~6 of the algorithm listings in~\cite{McKilliam.etal.2008a} is incorrect; it should read $i = n + 1 - \protect\lfloor (n + 1) (z_t + 0.5) \protect\rfloor$.
} in~\cite{McKilliam.etal.2008a}, as it gives the best known scaling (linear) with the lattice dimension $n$.
Essentially, the algorithm maps points in the space of $\Ans$ to points in the space of $\Zn[n+1]$, finds the nearest point in $\Zn[n+1]$ using Eq.~\eqref{eq:nearest-point-Zn}, then efficiently determines which point in $\Ans$ this point corresponds to.
The algorithm takes as input an $(n+1)$-dimensional vector $(0, \mat T\inv \vec\lambda)$ and returns an $(n+1)$-dimensional vector $\vec k\pr \in \bbZ^{n+1}$, from which the nearest template is given by
\begin{gather}
\label{eq:nearest-point-Ans-k}
\vec k = ( k\pr_1 - k\pr_0, k\pr_2 - k\pr_0, \dots, k\pr_n - k\pr_0 ) \,, \\
\label{eq:nearest-point-Ans}
\vec\lambda\Utmpl(\vec\lambda)|_{\Ans} = \mat T \vec k \,.
\end{gather}

We also would like to have an efficient lookup table from every template $\vec\lambda\Utmpl \in \{\vec\lambda\Utmpl\}$ to a unique index $j(\vec\lambda\Utmpl) = 0, 1, \dots, \calN - 1$, where $\calN$ is the number of templates in $\{\vec\lambda\Utmpl\}$.
For example, $j(\vec\lambda\Utmpl)$ might index an array of the $\calN$ values of the $\calF$-statistic computed at each template.
The lookup table would also be needed when implementing a semicoherent search, as discussed in Section~\ref{sec:discussion}.

\begin{figure}
\centering
\includegraphics[width=\linewidth]{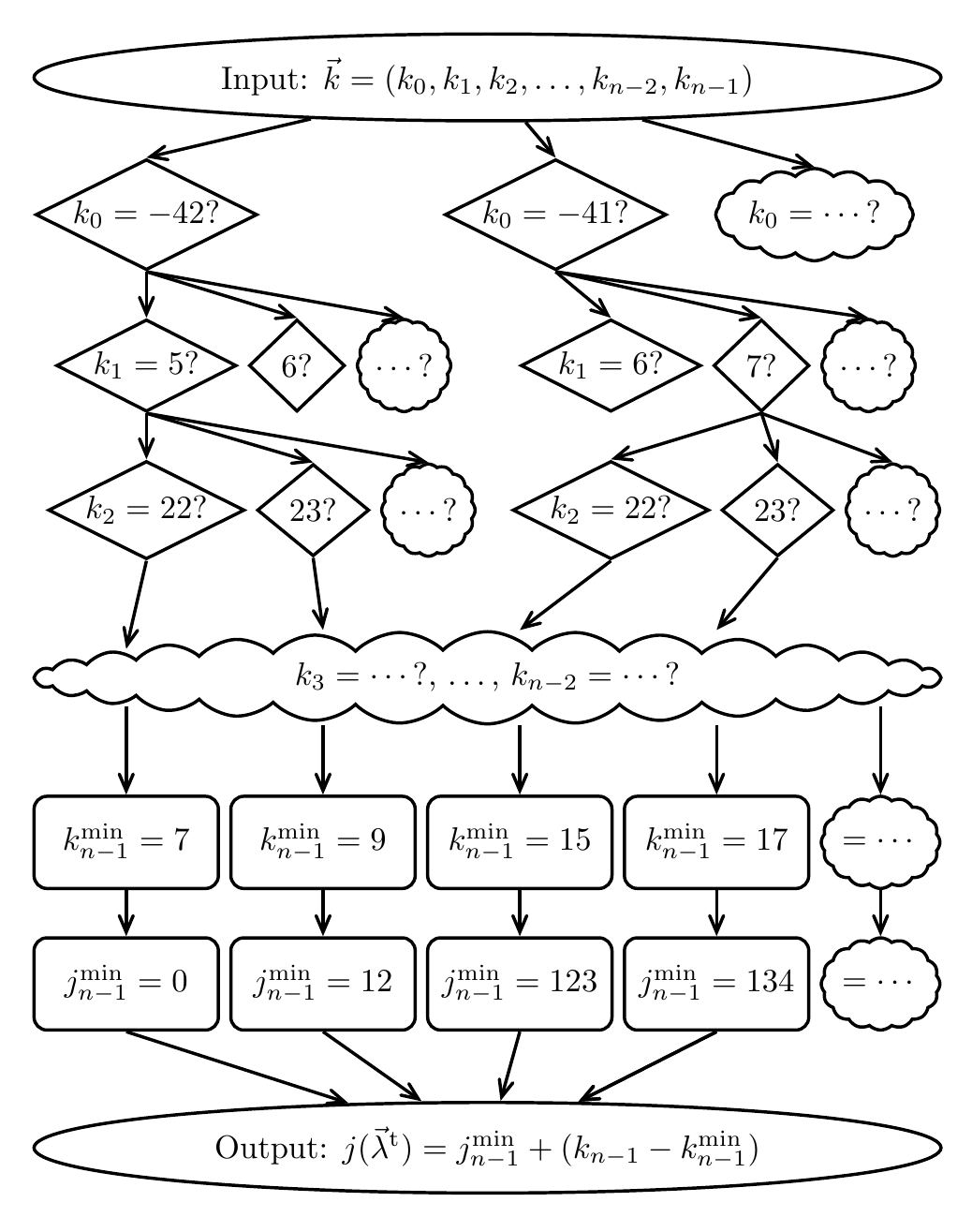}
\caption{\label{fig:lookup_trie_diagram}
Illustration of the $j(\vec\lambda\Utmpl)$ lookup trie described in Section~\ref{sec:near-templ-find-idx}.
The trie is traversed top to bottom.
Ovals denote the input and output nodes of the trie.
Diamonds denote nodes where a branch is taken.
Rectangles denote nodes where a stored value is retrieved.
Cloud shapes denote nodes omitted for brevity.
}
\end{figure}

We implement the $j(\vec\lambda\Utmpl)$ lookup table using a \emph{trie}, a tree data structure also known as a digital tree~\citep[e.g.][]{Knuth.vol3.1998}.
A trie has the advantage, compared to a hash table, of a constant lookup time of order $n$, and no possibility of key collisions which may degrade performance.
Its operation is illustrated in Fig.~\ref{fig:lookup_trie_diagram}.
The trie takes the vector $\vec k$ given by Eq.~\eqref{eq:nearest-point-Ans-k} as input.
First, $k_0$ determines which edge to follow from the input node; next, $k_1$ determines which edge to follow from the chosen $k_0$ node; $k_2$ then determines which edge to follow from the chosen $k_1$ node; and so on up to $k_{n-2}$.
From the chosen $k_{n-2}$ node, values for the $k_{n-1}$ lower bound $k\Umin_{n-1}$ and its index $j\Umin_{n-1}$ are retrieved.
Together with $k_{n-1}$, the index $j(\vec\lambda\Utmpl)$ is then given by
\begin{equation}
\label{eq:lookup-table-index}
j(\vec\lambda\Utmpl) = j\Umin_{n-1} + (k_{n-1} - k\Umin_{n-1}) \,.
\end{equation}
For example, for the example trie in Fig.~\ref{fig:lookup_trie_diagram}, a template with $\vec k = (-42, 5, 23, \cdots, 11)$ would have $k\Umin_{n-1} = 9$, $j\Umin_{n-1} = 12$, and hence $j(\vec\lambda\Utmpl) = 12 + (11 - 9) = 14$.
The lookup trie is constructed by generating each template in the bank, and filling the trie with the $k_0, \dots, k_{n-2}$ of each template, and the $k\Umin_{n-1}$ and $j\Umin_{n-1}$ of templates in the $(n-1)$th dimension.

\subsection{Lattice template placement testing}\label{sec:latt-templ-plac-test}

\begin{table*}
\caption{\label{tab:tmpl_placement_stats}
Input parameters to tests of the lattice template placement algorithm (Section~\ref{sec:latt-templ-plac}); properties of the template banks generated by the tests, averaged over $\Delta t\ustart$ at fixed $T$; and errors in estimates of the number of templates (Section~\ref{sec:number-templates}).
}
\begin{tabular*}{\linewidth}{l@{\extracolsep{\fill}}r@{\extracolsep{0pt}}l@{\extracolsep{2\tabcolsep}}r@{\extracolsep{0pt}}l@{\extracolsep{2\tabcolsep}}r@{\extracolsep{0pt}}l@{\extracolsep{2\tabcolsep}}r@{\extracolsep{0pt}}l@{\extracolsep{\fill}}r@{\extracolsep{0pt}}l@{\extracolsep{2\tabcolsep}}r@{\extracolsep{0pt}}l@{\extracolsep{2\tabcolsep}}r@{\extracolsep{0pt}}l@{\extracolsep{2\tabcolsep}}r@{\extracolsep{0pt}}l@{\extracolsep{\fill}}r@{\extracolsep{0pt}}l@{\extracolsep{2\tabcolsep}}r@{\extracolsep{0pt}}l@{\extracolsep{2\tabcolsep}}r@{\extracolsep{0pt}}l@{\extracolsep{\fill}}r@{\extracolsep{0pt}}l@{\extracolsep{2\tabcolsep}}r@{\extracolsep{0pt}}l@{\extracolsep{2\tabcolsep}}r@{\extracolsep{0pt}}l@{\extracolsep{2\tabcolsep}}}
\hline\hline
Quantity & \multicolumn{28}{c}{Value for parameter space} \\
 & \multicolumn{8}{c}{A$_1$} & \multicolumn{8}{c}{A$_2$} & \multicolumn{6}{c}{B$_1$} & \multicolumn{6}{c}{B$_2$} \\
\hline
Starting frequency / Hz & \multicolumn{8}{c}{$\nu = 100$} & \multicolumn{8}{c}{$\nu = 500$} & \multicolumn{6}{c}{$f = 100$} & \multicolumn{6}{c}{$f = 500$} \\
Frequency band / Hz & \multicolumn{8}{c}{$\Delta\nu = 10^{-6}$} & \multicolumn{8}{c}{$\Delta\nu = 10^{-6}$} & \multicolumn{6}{c}{$\Delta f = 10^{-6}$} & \multicolumn{6}{c}{$\Delta f = 10^{-6}$} \\
First spindown band / Hz & \multicolumn{8}{c}{$\Delta\dot\nu = 10^{-9}$} & \multicolumn{8}{c}{$\Delta\dot\nu = 10^{-9}$} & \multicolumn{6}{c}{$\Delta\dot f = 10^{-9}$} & \multicolumn{6}{c}{$\Delta\dot f = 10^{-9}$} \\
Maximum mismatch & \multicolumn{8}{c}{ $\mu\umax = 0.3$} & \multicolumn{8}{c}{ $\mu\umax = 0.6$} & \multicolumn{6}{c}{ $\mu\umax = 0.3$} & \multicolumn{6}{c}{ $\mu\umax = 0.6$} \\
Time span $T$ / days & \multicolumn{2}{c}{1} & \multicolumn{2}{c}{3} & \multicolumn{2}{c}{9} & \multicolumn{2}{c}{27} & \multicolumn{2}{c}{1} & \multicolumn{2}{c}{3} & \multicolumn{2}{c}{9} & \multicolumn{2}{c}{27} & \multicolumn{2}{c}{1} & \multicolumn{2}{c}{3} & \multicolumn{2}{c}{9} & \multicolumn{2}{c}{1} & \multicolumn{2}{c}{3} & \multicolumn{2}{c}{9} \\
\hline
$\log_{10}(\text{Number of templates~} \calN)$ & $6$ & $.3$ & $6$ & $.8$ & $7$ & $.2$ & $8$ & $.7$ & $7$ & $.3$ & $7$ & $.7$ & $8$ & $.1$ & $9$ & $.6$ & $7$ & $.8$ & $8$ & $.2$ & $9$ & $.4$ & $8$ & $.8$ & $9$ & $.1$ & $10$ & $.4$ \\
$\log_{10}(\text{Bulk templates} / \calN)$ & \multicolumn{2}{c}{$-3$} & \multicolumn{2}{c}{$-2$} & \multicolumn{2}{c}{$-1$} & $-0$ & $.8$ & \multicolumn{2}{c}{$-3$} & \multicolumn{2}{c}{$-2$} & \multicolumn{2}{c}{$-1$} & $-0$ & $.9$ & \multicolumn{2}{c}{$-5$} & \multicolumn{2}{c}{$-4$} & \multicolumn{2}{c}{$-4$} & \multicolumn{2}{c}{$-5$} & \multicolumn{2}{c}{$-4$} & \multicolumn{2}{c}{$-4$} \\
$\log_{10}(\text{Missed test points} / \calN)$ & \multicolumn{2}{c}{$-\infty$} & \multicolumn{2}{c}{$-\infty$} & \multicolumn{2}{c}{$-\infty$} & \multicolumn{2}{c}{$-\infty$} & \multicolumn{2}{c}{$-\infty$} & \multicolumn{2}{c}{$-\infty$} & \multicolumn{2}{c}{$-\infty$} & \multicolumn{2}{c}{$-\infty$} & \multicolumn{2}{c}{$-\infty$} & \multicolumn{2}{c}{$-16$} & \multicolumn{2}{c}{$-17$} & \multicolumn{2}{c}{$-\infty$} & \multicolumn{2}{c}{$-14$} & \multicolumn{2}{c}{$-15$} \\
Mean reduced supersky $\avg{\mu\urss}$ & $0$ & $.16$ & $0$ & $.16$ & $0$ & $.16$ & $0$ & $.16$ & $0$ & $.31$ & $0$ & $.31$ & $0$ & $.31$ & $0$ & $.31$ & $0$ & $.16$ & $0$ & $.16$ & $0$ & $.16$ & $0$ & $.31$ & $0$ & $.31$ & $0$ & $.31$ \\
Mean supersky $\avg{\mu\uss}$ & $0$ & $.15$ & $0$ & $.15$ & $0$ & $.16$ & $0$ & $.16$ & $0$ & $.31$ & $0$ & $.31$ & $0$ & $.31$ & $0$ & $.32$ & $0$ & $.15$ & $0$ & $.15$ & $0$ & $.16$ & $0$ & $.31$ & $0$ & $.31$ & $0$ & $.31$ \\
\hline
$\langle |\calN\uest(\text{num.~} \mat g\urss)-\calN|/\calN \rangle$ / \% & $1$ & $.8$ & $1$ & $.3$ & $2$ & $.6$ & $5$ & $.6$ & $4$ & $.3$ & $2$ & $.4$ & $3$ & $.8$ & $2$ & $.3$ & $8$ & $.1$ & $1$ & $.5$ & $1$ & $.8$ & $3$ & $.4$ & $0$ & $.3$ & $0$ & $.5$ \\
$\max |\calN\uest(\text{num.~} \mat g\urss)-\calN|/\calN$ / \% & $2$ & $.3$ & $4$ & $.8$ & $4$ & $.7$ & $7$ & $.5$ & $6$ & $.8$ & $3$ & $.6$ & $16$ & $.5$ & $2$ & $.9$ & $8$ & $.4$ & $1$ & $.5$ & $2$ & $.1$ & $3$ & $.5$ & $0$ & $.3$ & $0$ & $.6$ \\
$\langle |\calN\uest(\text{appx.~} \mat g\urss)-\calN|/\calN \rangle$ / \% & $1$ & $.7$ & $4$ & $.1$ & $13$ & $.8$ & $6$ & $.0$ & $3$ & $.4$ & $4$ & $.9$ & $17$ & $.8$ & $3$ & $.8$ & $78$ & $.9$ & $44$ & $.7$ & $48$ & $.8$ & $80$ & $.5$ & $45$ & $.5$ & $49$ & $.1$ \\
$\max |\calN\uest(\text{appx.~} \mat g\urss)-\calN|/\calN$ / \% & $2$ & $.2$ & $7$ & $.7$ & $17$ & $.0$ & $10$ & $.0$ & $6$ & $.0$ & $6$ & $.1$ & $33$ & $.7$ & $6$ & $.1$ & $88$ & $.5$ & $68$ & $.8$ & $73$ & $.4$ & $89$ & $.4$ & $69$ & $.2$ & $73$ & $.8$ \\
\hline\hline
\end{tabular*}
\end{table*}

This section presents tests performed on lattice template banks generated by the algorithm described in the previous sections.
Details of input parameters to the tests are summarized in Table~\ref{tab:tmpl_placement_stats}.

Template banks are generated using four types of parameter spaces, labeled A$_1$, A$_2$, B$_1$, and~B$_2$.
The maximum mismatches $\mu\umax$ used for each type, and the list of time spans $T$ for which templates banks of each type are generated, are given in Table~\ref{tab:tmpl_placement_stats}.
All types are also generated at five values of start time $\Delta t\ustart = t\uref \pm \{0, 90, 180\}$~days, where $t\uref$ is given in Section~\ref{sec:fixed-reference-time}.

Parameter spaces~A$_X$ ($X = 1, 2$) cover the whole sky, a fixed band in reduced supersky frequency $\nu$ of width $10^{-6}$~Hz, and a fixed band in reduced supersky spindown $\dot \nu$ of $[-10^{-9},0]$~Hz~s$\inv$.
The B$_X$~parameter spaces cover the whole sky, a fixed band in physical frequency $f$ of width $10^{-6}$~Hz, and a fixed band in physical spindown $\dot f$ of $[-10^{-9},0]$~Hz~s$\inv$.
The frequency bands start at 100~Hz for A$_1$ and~B$_1$, and at 500~Hz for A$_2$ and~B$_2$.
The bandwidths are limited by the computational cost of generating the template banks and calculating the mismatches to each test point.

For each template $\vec\lambda\Utmpl$ in each generated template bank, $\sim 10$~test points $\vec\lambda$ are randomly drawn from the parameter space, and the nearest template $\vec\lambda\Utmpl(\vec\lambda)$ to each test point is found as described in Section~\ref{sec:near-templ-find-idx}.
The mismatches $\mu\urss$ and $\mu\uss$ between each test point and their nearest templates are computed via Eq.~\eqref{eq:metric-mismatch} for both reduced supersky and supersky metrics.

Table~\ref{tab:tmpl_placement_stats} lists, for each parameter space and each $T$, the following properties computed by the tests: the number of templates; the fraction of ``bulk'' templates, i.e.\ excluding extra templates needed to cover the boundaries (see Fig.~\ref{fig:boundary_covering_diagram} and Section~\ref{sec:stairc-bound-templ}); the fraction of missed test points, i.e.\ where $\mu\urss > \mu\umax$; and the means of the mismatch distributions of $\mu\urss$ and $\mu\uss$.
These quantities are averaged over the five start times $\Delta t\ustart$.

Due to the limited frequency and spindown bandwidths, the template banks are dominated by boundary templates, as evidenced by the relatively small fractions of bulk templates (Table~\ref{tab:tmpl_placement_stats}).
Since the simulations in Section~\ref{sec:refin-numer-simul} effectively test the properties of an infinite template bank, i.e.\ without boundaries, it is complementary that the tests presented here test the properties of template banks dominated by boundaries.

The fractions of missed test points (Table~\ref{tab:tmpl_placement_stats}) indicate ``holes'' in the template bank, i.e.\ regions of parameter space not covered by templates to within the desired maximum mismatch.
No holes were found in parameter spaces~A$_X$, i.e.\ no test points were missed, and only very small holes were found in B$_X$, the number of missed test points per template being $\lesssim 10^{-14}$.

\begin{figure*}
\centering
\subfloat[]{\includegraphics[width=0.49\linewidth]{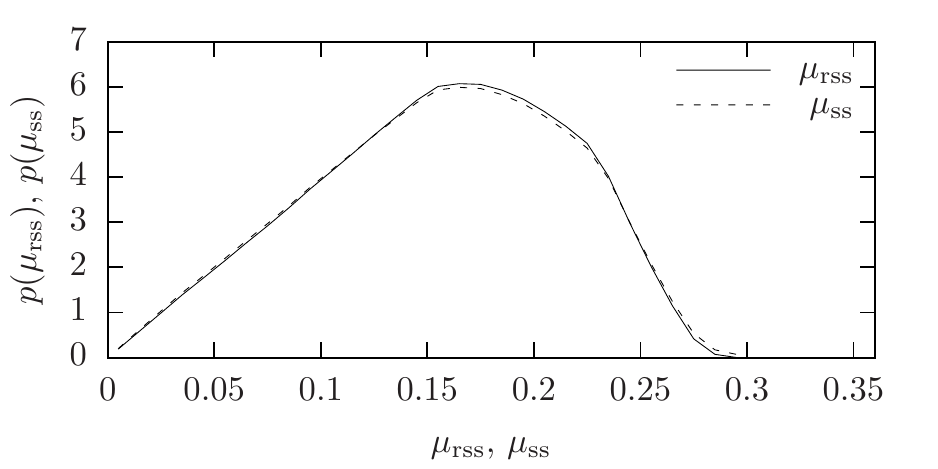}\label{fig:tmpl_mu_hgrm_nu100_f1dot_H1L1_mm0p3_Ans}}
\subfloat[]{\includegraphics[width=0.49\linewidth]{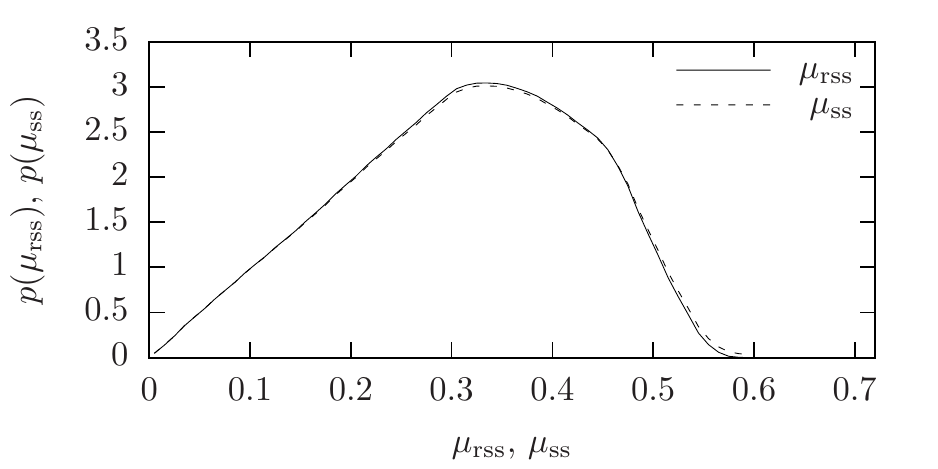}\label{fig:tmpl_mu_hgrm_nu500_f1dot_H1L1_mm0p6_Ans}}\\
\subfloat[]{\includegraphics[width=0.49\linewidth]{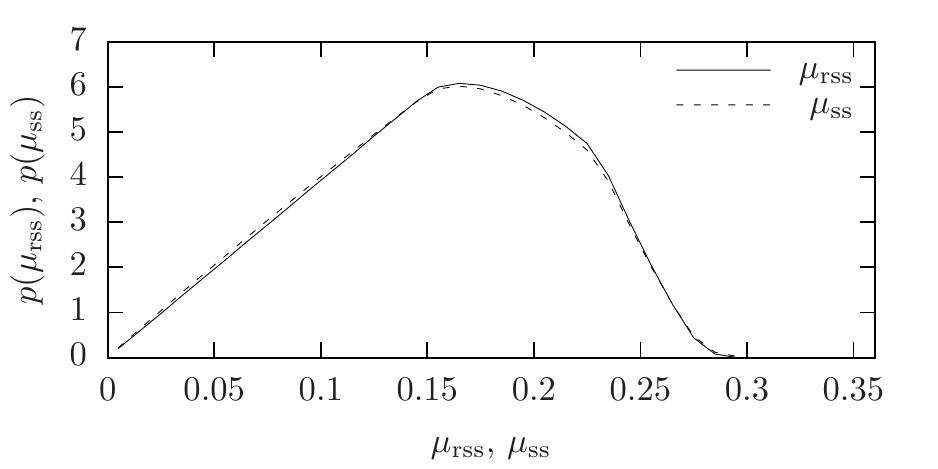}\label{fig:tmpl_mu_hgrm_f100_f1dot_H1L1_mm0p3_Ans}}
\subfloat[]{\includegraphics[width=0.49\linewidth]{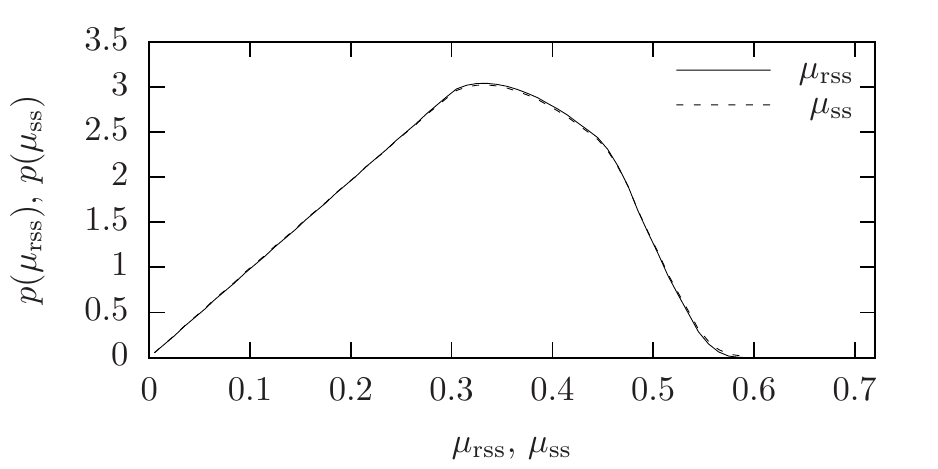}\label{fig:tmpl_mu_hgrm_f500_f1dot_H1L1_mm0p6_Ans}}
\caption{\label{fig:tmpl_mu_hgrm}
Histograms of mismatches in $\mu\urss$ and $\mu\uss$, generated by the lattice template placement algorithm detailed in Section~\ref{sec:latt-templ-plac}, and averaged over all values of $T$ and $\Delta t\ustart$.
The plots correspond to the parameter spaces \protect\subref{fig:tmpl_mu_hgrm_nu100_f1dot_H1L1_mm0p3_Ans}~A$_1$, \protect\subref{fig:tmpl_mu_hgrm_nu500_f1dot_H1L1_mm0p6_Ans}~A$_2$, \protect\subref{fig:tmpl_mu_hgrm_f100_f1dot_H1L1_mm0p3_Ans}~B$_1$, and \protect\subref{fig:tmpl_mu_hgrm_f500_f1dot_H1L1_mm0p6_Ans}~B$_2$ from Table~\ref{tab:tmpl_placement_stats}.
An $\Ans$ lattice is used to place templates with $\mu\urss \le 0.3$ (left column) and $\mu\urss \le 0.6$ (right column).
}
\end{figure*}

The mean mismatches $\avg{\mu\urss}$ and $\avg{\mu\uss}$, shown in Table~\ref{tab:tmpl_placement_stats}, are similar for the reduced supersky and supersky metrics respectively.
Figure~\ref{fig:tmpl_mu_hgrm} plots the averaged mismatch distributions of $\mu\urss$ and $\mu\uss$ for each parameter space, which also show close agreement.
While the $\calF$-statistic mismatch was not computed in these tests, the similarity of the histograms in Fig.~\ref{fig:tmpl_mu_hgrm} to those in Fig.~\ref{fig:mu_hgrm_f1dot_H1L1_mm0p3_Ans} leads us to expect (were it to be computed) a $\calF$-statistic mismatch histogram similar to that in Fig.~\ref{fig:mu_hgrm_f1dot_H1L1_mm0p3_Ans}.

\subsection{The ``staircase'' boundary template issue}\label{sec:stairc-bound-templ}

This section describes an issue that is encountered when covering the boundaries of certain parameter spaces, in particular the parameter spaces~B$_X$ of the previous section.
We call it the ``staircase'' issue.
It arises because the boundaries of the parameter space, which are represented per Eq.~\eqref{eq:param-space-lambda} by continuous functions $\lambda\Umin(\dots)$ and $\lambda\Umax(\dots)$, are being covered by a bank of discrete templates $\{\lambda\Utmpl\}$.
If, in parameter-space dimension $i$, the $\lambda\Umin_i(\dots)$ and $\lambda\Umax_i(\dots)$ functions change significantly on the scale of a single template, e.g.\ the width of its metric ellipse bounding box $\beta_i$, regions of the parameter space boundary in dimension $i$ may not be covered.

\begin{figure}
\includegraphics[width=\linewidth]{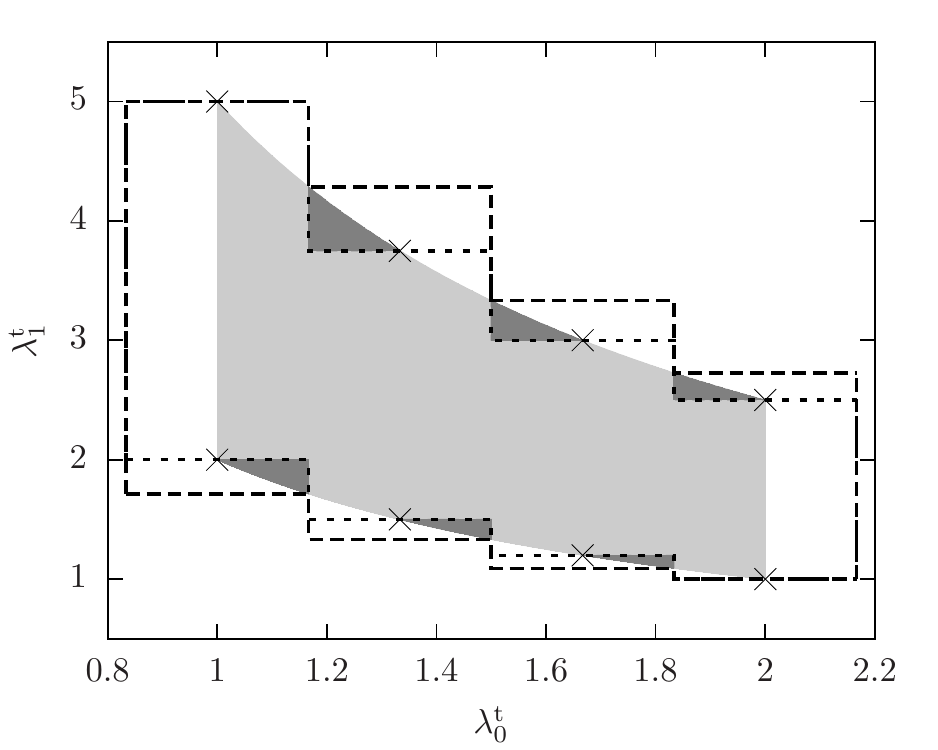}
\caption{\label{fig:parameter_space_staircase}
Illustration of the staircase boundary template issue, using the example parameter space of Fig.~\ref{fig:parameter_space_example}.
The parameter space to be covered is light gray.
The short-dashed outline shows the area covered by the template bank, if the bounds
$\lambda\Umin_1(\lambda\Utmpl_0)$ and $\lambda\Umax_1(\lambda\Utmpl_0)$ are computed exactly at the boundary templates (crosses).
Regions of the parameter space not covered by this template bank are dark gray.
The long-dashed outline shows the area covered if instead the bounds are extremized over the bounding box of each boundary template.
}
\end{figure}

Figure~\ref{fig:parameter_space_staircase} illustrates an example of the staircase issue.
The example parameter space of Fig.~\ref{fig:parameter_space_example}, shaded light gray, is covered with a template bank which is deliberately coarse in the $\lambda\Utmpl_0$ coordinate, comprising only four columns of templates; the boundary templates of these columns are plotted as crosses.
For simplicity, we take the region covered by this template bank to be the union of the bounding boxes around each template; this gives the short-dashed, staircase-shaped outline.
(For the purpose of this illustration, we ignore the extra boundary templates illustrated in Fig.~\ref{fig:boundary_covering_diagram}.)

It is clear that there are regions of the parameter space, shaded dark gray, not covered by this template bank.
It is also be clear that the uncovered regions exist because the short-dashed outline is a poor approximation to the true parameter-space shape.
A much denser template bank in $\lambda\Utmpl_0$, so that the bound functions $\lambda\Umin_1(\lambda\Utmpl_0)$ and $\lambda\Umax_1(\lambda\Utmpl_0)$ are better approximated, would not suffer from this issue.
By analogy, imagine trying to approximate the area under a curve $f(x)$ by a series of rectangles of fixed width $\delta x$ and centered at points $x_j$, i.e.\ $\int f(x) dx \approx \sum_j f(x_j) \delta x$.
We would expect a poor approximation if $f(x_j + \delta x) \not\approx f(x_j)$ for many $x_j$.
Conversely, one would expect a good approximation if $f(x_j + \delta x) \approx f(x_j)$ for most $x_j$, which may be achieved simply by reducing $\delta x$ and adding more points $x_j$.

In the case of template placement, however, it would be undesirable to increase the template bank density, since this would inflate the computational cost of the search.
We can, instead, improve how well the region covered by the template bank approximates the true shape of the parameter space.
Instead of calculating the bound functions $\lambda\Umin_i(\cdots)$ and $\lambda\Umax_i(\cdots)$ solely at a boundary template, we can find their extrema over the region covered by that template, e.g.\ its bounding box.
This ensures that the template bank extends far enough to cover any variations in the boundaries that might occur between neighboring boundary templates.

For example, in Fig.~\ref{fig:parameter_space_staircase}, we can replace the functions $\lambda\Umin_1(\lambda\Utmpl_0)$ and $\lambda\Umax_1(\lambda\Utmpl_0)$ with the following:
\begin{subequations}
\label{eqs:extreme-bounds-example}
\begin{align}
\lambda\Umin_1(\lambda_0) &\rightarrow \min_{|\delta\lambda_0| \le 0.5\beta_0} \lambda\Umin_1(\lambda_0 + \delta\lambda_0) \,, \\
\lambda\Umax_1(\lambda_0) &\rightarrow \max_{|\delta\lambda_0| \le 0.5\beta_0} \lambda\Umax_1(\lambda_0 + \delta\lambda_0) \,,
\end{align}
\end{subequations}
where $\beta_0$ is the bounding box width in the $\lambda_0$ coordinate.
Doing so gives the long-dashed outline in Fig.~\ref{fig:parameter_space_staircase}, which now completely covers the parameter space.

The generalization of Eqs.~\eqref{eqs:extreme-bounds-example} to any parameter-space bound function is, e.g.\ for $\lambda\Umin_i(\cdots)$:
\begin{multline}
\label{eq:extreme-bounds}
\lambda\Umin_i(\lambda_0, \cdots, \lambda_{i-1}) \rightarrow \\
\min_{\substack{|\delta\lambda_0| \le 0.5\beta_0 \\ \cdots \\ |\delta\lambda_{i-1}| \le 0.5\beta_{i-1}}}
\lambda\Umin_1(\lambda_0 + \delta\lambda_0, \cdots, \lambda_{i-1} + \delta\lambda_{i-1}) \,.
\end{multline}
If it is known that the bound functions are monotonic over the extent of the bounding box, the extrema can only occur at the vertices of the bounding box.
It is then sufficient to extremize the bounds only over the finite set of vertices, e.g.:
\begin{multline}
\label{eq:extreme-bounds-finite}
\lambda\Umin_i(\lambda_0, \cdots, \lambda_{i-1}) \rightarrow \\
\min \lambda\Umin_1(\lambda_0 \pm 0.5\beta_0, \cdots, \lambda_{i-1} \pm 0.5\beta_{i-1}) \,.
\end{multline}

The staircase issue appears when covering the parameter spaces B$_X$ of Section~\ref{sec:latt-templ-plac-test}.
While these parameter spaces cover fixed bands in physical frequency/spindown $\ndot f$,
the parameter-space bounds are specified as functions of the reduced supersky frequencies/spindowns $\ndot \nu$.
The two coordinate systems are related by~(see \PaperI)
\begin{equation}
\ndot \nu = \ndot f + \vec\Delta^s \cdot \vec n \,,
\end{equation}
where the $\vec\Delta^s$ are offset vectors found when computing the reduced supersky metric.
When covering a fixed band in $\ndot f$, the bounds on $\ndot \nu$, e.g.\ $\ndot \nu{}\Umin(\vec n)$, are therefore functions of sky position $\vec n$.

The difference in $\ndot \nu{}\Umin(\vec n)$ between neighboring sky positions $\vec n_1$ and $\vec n_2$ is
\begin{equation}
\label{eq:nu-min-difference}
\begin{split}
|\ndot \nu{}\Umin(\vec n_1) - \ndot \nu{}\Umin(\vec n_2)| &= |\vec\Delta^s \cdot (\vec n_1 - \vec n_2)| \\
&\sim |\Delta\ua^s| \beta\ua + |\Delta\ub^s| \beta\ub \,,
\end{split}
\end{equation}
where $\vec\beta$ is the metric ellipse bounding box given by Eq.~\eqref{eq:bounding-box}.
Depending on the density of sky templates, this difference can be much larger than the extent of the bounding box in $\ndot \nu$.
For example, when $T \sim 1$~day, $|\nu\Umin(\vec n_1) - \nu\Umin(\vec n_2)| \sim 10^{-3}$~Hz, whereas $\beta_{\nu} \sim 10^{-5}$~Hz.
In short, the bounds on $\nu$ are changing on a much larger scale ($\sim 10^{-3}$~Hz) than the extent of a single template ($\sim 10^{-5}$~Hz); this is precisely the conditions where the staircase issue becomes important.

At the time the tests in Section~\ref{sec:latt-templ-plac-test} were performed, the solution to the staircase issue outlined above [i.e.\ Eq.~\eqref{eq:extreme-bounds}] was not realized.
Instead, an empirical solution was used, where the bounds on $\ndot\nu$ were simply extended by $|\Delta\ua^s| \beta\ua + |\Delta\ub^s| \beta\ub$ [Eq.~\eqref{eq:nu-min-difference}].
This solution results in only a small fraction of missed test points in the B$_X$ parameter spaces (Table~\ref{tab:tmpl_placement_stats}).
The solution suggested by Eq.~\eqref{eq:extreme-bounds-finite} will however be used in future implementations.

\section{Number of templates}\label{sec:number-templates}

Having confirmed, in the previous section, that lattice template placement using the reduced supersky metric can be successfully implemented, we now investigate the number of templates required to cover an all-sky parameter space.
Section~\ref{sec:count-estim-numb} considers how to accurately estimate the number of templates; this is needed in order to determine the most sensitive semicoherent search setup~\cite[e.g.][]{Prix.Shaltev.2012a}.
Section~\ref{sec:scal-numb-templ} examines the scaling of the number of templates with time span $T$, which determines the relationship between search sensitivity and computational cost.

\subsection{Counted and estimated number of templates}\label{sec:count-estim-numb}

The number of templates required to cover an $n$-dimensional parameter space $\calP$ with constant metric $\mat g$, using a lattice template bank with maximum mismatch $\mu\umax$, is estimated by~\cite{Prix.2007b,Jaranowski.Krolak.2012}
\begin{equation}
\label{eq:Npred}
\calN\uest = \theta \mu\umax^{-n/2} \sqrt{\det \mat g} \int_{\calP \cup \partial\calP} d\vec\lambda \,,
\end{equation}
where $\theta$ is the normalized thickness of the lattice being used (see Section~\ref{sec:lattices}).
The integral gives the volume, with respect to the parameters $\vec\lambda$, of both $\calP$ and its boundary $\partial\calP$, to account for the extra boundary templates discussed in Sections~\ref{sec:param-space-repr} and~\ref{sec:stairc-bound-templ}.

The determinant of the reduced supersky metric may be written as
\begin{equation}
\label{eq:det-g-urss}
\det \mat g\urss = \det \mat g\urssnn \cdot \det \mat g\urssnunu \,,
\end{equation}
where $\mat g\urssnn$ is the sky--sky block and $\mat g\urssnunu$ the frequency--frequency block of $\mat g\urss$, the off-diagonal sky--frequency blocks being zero.
The matrix $\mat g\urssnunu$ is identical to the physical frequency/spindown metric, for which analytic expressions exist~\cite[e.g][]{Whitbeck.2006a,Wette.etal.2008a,Prix.2013}.
Its determinant is therefore known exactly, e.g.:
\begin{equation}
\label{eq:det-g-urss-nunu}
\det \mat g\urssnunu = \begin{cases}
\frac{ \pi^{4} T^{6} }{ 540 } \,, & s\umax = 1 \,, \\
\frac{ \pi^{6} T^{12} }{ 13608000 } \,, & s\umax = 2 \,,
\end{cases}
\end{equation}
where $s\umax$ is the maximum number of spindowns.
The matrix $\mat g\urssnn$ is diagonal, and its determinant is therefore the product of its diagonal elements $g\unana$ and $g\unbnb$.
In turn, these are reasonably well-approximated by functions of $T$ only, ignoring the weaker dependence on $t\ustart$ due to the Earth's noncircular orbital motion.
Analytic expressions approximating $g\unana$ and $g\unbnb$ are given by Eqs.~\eqref{eqs:metric-fit-gnabnab-eqn} in the Appendix.

Table~\ref{tab:tmpl_placement_stats} shows errors $|\calN\uest-\calN|/\calN$ between the number of templates $\calN$ counted in the template banks generated in Section~\ref{sec:latt-templ-plac}, and the number $\calN\uest$ estimated using Eq.~\eqref{eq:Npred}.
The $\mat g\urss$ used in Eq.~\eqref{eq:Npred} is either computed numerically, or approximated using Eqs.~\eqref{eq:det-g-urss}, \eqref{eq:det-g-urss-nunu}, and \eqref{eqs:metric-fit-gnabnab-eqn}.
Using the numerically-computed $\mat g\urss$, errors are a few percent on average, and are limited to $\lesssim 17\%$ over the four types of parameter spaces.
This confirms that the behavior of the algorithm of Section~\ref{sec:latt-templ-plac} conforms to that expected by Eq.~\eqref{eq:Npred}.
The approximate $\mat g\urss$ also leads to reasonable errors of $\lesssim 34\%$, except for parameter spaces B$_X$, where they are between $44\%$ and $90\%$.
This is due to using typical values for the elements of the offset vectors $\vec\Delta^s$ when estimating the number of extra staircase boundary templates: $|\Delta\ua^0| \sim |\Delta\ub^0| \sim 5{\times}10^{-5} f\umax$ and $|\Delta\ua^1| \sim |\Delta\ub^1| \sim 10^{-11} f\umax$.
Nevertheless, we imagine that the approximate method may still be useful for rapid order-of-magnitude template counting, e.g.\ during search setup optimization~\citep[e.g.][]{Prix.Shaltev.2012a}.

\begin{figure}
\centering
\subfloat[]{\includegraphics[width=\linewidth]{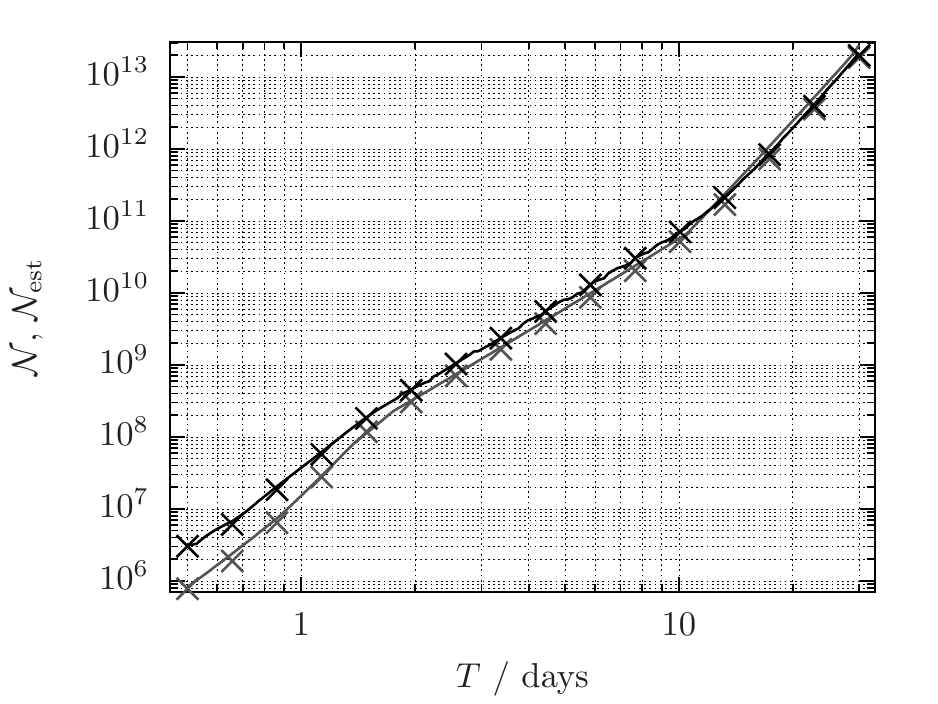}\label{fig:number_of_tmpls_f1dot}}\\
\subfloat[]{\includegraphics[width=\linewidth]{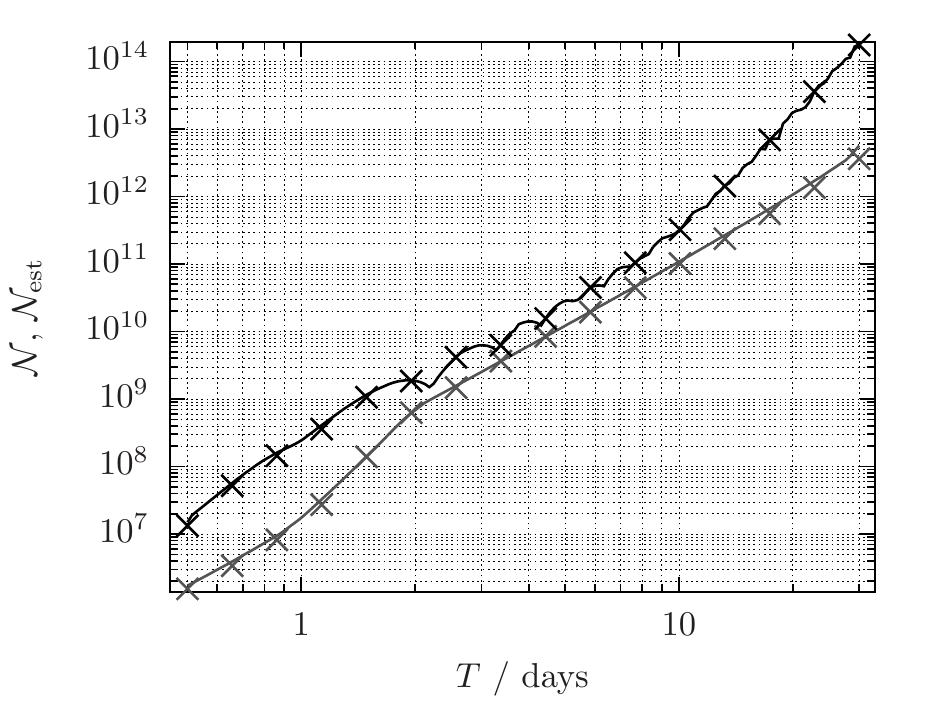}\label{fig:number_of_tmpls_f2dot}}
\caption{\label{fig:number_of_tmpls}
Number of templates required to search a coherent segment of time span $T$, centered on the reference time $t\uref$ of Section~\protect\ref{sec:fixed-reference-time}, over the following parameter spaces, using an $\Ans$ lattice template bank with $\mu\umax = 0.3$:
\protect\subref{fig:number_of_tmpls_f1dot} the whole sky, a fixed band in $f$ of $[100, 100.01]$~Hz, and a fixed band in $\dot f$ of $[-10^{-8},0]$~Hz~s$\inv$;
\protect\subref{fig:number_of_tmpls_f2dot} as above, with an additional fixed band in $\ddot f$ of $[0,10^{-18}]$~Hz~s$\inv[2]$.
Crosses denote the number $\calN$ counted by the algorithm of Section~\protect\ref{sec:latt-templ-plac}; lines denote the number $\calN\uest$ estimated by the procedures described in Section~\protect\ref{sec:number-templates}.
Black/gray denotes the template counts including/excluding the extra boundary templates discussed in Section~\protect\ref{sec:stairc-bound-templ}.
}
\end{figure}

Figure~\ref{fig:number_of_tmpls} plots the number of templates, as functions of $T$, required to search a more realistic parameter space  than those used for testing in Section~\ref{sec:latt-templ-plac}: the whole sky, a frequency band of $10^{-2}$~Hz at $f = 100$~Hz, and bands of widths $10^{-8}$~Hz~s$\inv$ in first spindown and $10^{-18}$~Hz~s$\inv[2]$ in second spindown.
(In particular, the frequency band is more typical of an Einstein@Home ``work unit''~\cite{Aasi.etal.2013a})
The number of templates is counted by the algorithm of Section~\ref{sec:latt-templ-plac}, and estimated using Eq.~\eqref{eq:Npred}.
Two estimates are performed: using the numerically-computed $\mat g\urss$ and taking into account the extra staircase boundary templates discussed in Section~\ref{sec:stairc-bound-templ}, and using the approximate $\mat g\urss$ and ignoring the extra staircase boundary templates.
Good agreement is seen between the number of counted templates and the various estimates.
For first spindown-only searches, the fraction of templates required to address the staircase issue steadily decreases with $T$; when second spindown is added, the fraction is larger, indicating that the second spindown band is more dominated by its boundaries.

\Citet{Brady.etal.1998a} predict the number of templates required to search the whole sky, frequencies up to 200~Hz, and spindowns $|\dot f| \le f / 10^3$~yr, motivated by a minimum ``spindown age'' of the gravitational-wave pulsar.
Their estimate, given by their Eqs.~(6.3)--(6.7) and~(6.9), used a different derivation of the parameter-space metric (based on loss of power, not loss of $\calF$-statistic) which is nonconstant; it is therefore an idealized estimate of the minimal number of templates achievable.
It assumed an hexagonal prism lattice (i.e.\ the composite lattice $\Ans[2] \otimes \Zn[n-2]$), and a maximum \emph{projected} mismatch of 0.3 (i.e.\ assuming mismatch has already been minimized over frequency).

\begin{figure}
\centering
\includegraphics[width=\linewidth]{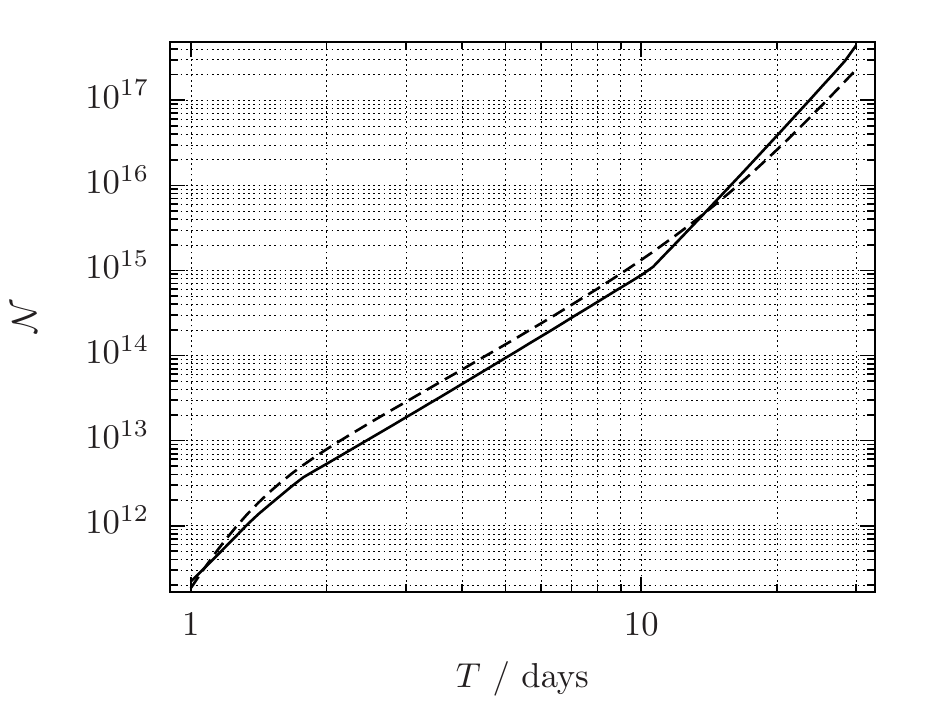}
\caption{\label{fig:brady_comparison}
Number of templates required to search a coherent segment of time span T over the whole sky, a fixed band in $f$ of $[0, 200]$~Hz, and frequency-dependent bands in $\dot f$ of $\pm f / (10^3~\mathrm{yr})$.
The estimate of~\citet{Brady.etal.1998a} (dashed line) is compared with an estimate following the procedure of Section~\ref{sec:number-templates} using the approximate $\mat g\urss$ (solid line).
}
\end{figure}

Figure~\ref{fig:brady_comparison} compares the estimate of~\cite{Brady.etal.1998a} to an estimate given by Eq.~\eqref{eq:Npred}, using the approximate $\det \mat g\urss$, an $\Ans$ lattice, and a maximum (total) mismatch of $\mu\umax = 0.3$.
The two estimates are in close agreement, differing by less than a factor of two.
Given the idealized assumptions of~\cite{Brady.etal.1998a}, it is encouraging to see close agreement with the template estimation described in this section, which in turn is in good agreement with the practical implementation of lattice template placement described in Section~\ref{sec:latt-templ-plac}.

\subsection{Scaling of number of templates with time span}\label{sec:scal-numb-templ}

The number of templates $\calN$ may be modeled, in the vicinity of a fixed time span $T_0$, by a power law in $T$: $\calN(T) |_{T=T_0} \approx T^p$, where the local power-law exponent is
\begin{equation}
p = \left. \frac{ d[\log \calN(T)] }{ d[\log T] } \right|_{T=T_0} \,.
\end{equation}
This exponent is used to relate the sensitivity of a search and its computational cost~\cite{Prix.Shaltev.2012a}.
The scaling of the number of frequency/spindown templates $\calN\ufreq$ is known exactly: the number of templates in each frequency/spindown coordinate $\ndot \nu$ scales as $T^{s+1}$, and hence $\calN\ufreq \propto T^{(s\umax + 1)(s\umax + 2)/2}$.
The scaling of the number of sky templates $\calN\usky$ is, however, less straightforward.
It is generally stated that $\calN\usky \propto T^q$ with $q \ge 2$~\cite{Brady.etal.1998a,Whitbeck.2006a,Abbott.etal.2007d,Prix.2007a,Prix.2009a,LSC.VC.2013a}; on the other hand,~\cite{Pletsch.2010a} reported $\calN\usky$ to be approximately constant once $T \gtrsim 2$~days.

\begin{figure}
\centering
\includegraphics[width=\linewidth]{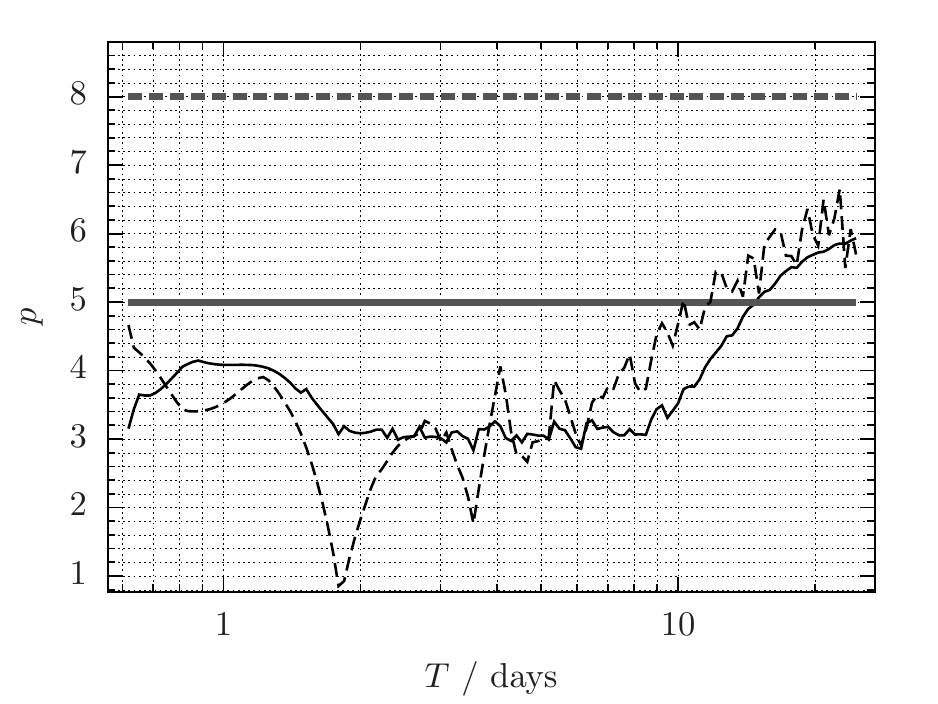}
\caption{\label{fig:number_of_tmpls_exponent}
Local power-law exponent $p$ of the number of templates plotted in Figs.~\protect\ref{fig:number_of_tmpls_f1dot}, with first spindown (solid line), and~\protect\ref{fig:number_of_tmpls_f2dot} (dashed line), with both first and second spindown.
The thick horizontal lines denote the expected $p$, assuming that $\calN\usky \propto T^2$ and $\calN\ufreq \propto T^3$ (solid) or $T^6$ (dashed).
}
\end{figure}

Figure~\ref{fig:number_of_tmpls_exponent} plots the local power-law exponent $p$ of the number of templates $\calN = \calN\usky \calN\ufreq$.
When including only first spindown, we find a similar scaling of $\calN\usky$ to that of~\cite{Pletsch.2010a}; $p \approx 3$ for $2 \lesssim T \lesssim 10$~days, which implies (since $\calN\ufreq$ is known to scale as $T^3$) that $\calN\usky$ is approximately constant over this period.
Once $T \gtrsim 10$~days, however, $\calN\usky$ scales with an increasing power $q \lesssim 3$.
When including both first and second spindown, the scaling of $\calN$ with $T$ becomes more complicated, which is likely due to the Earth's noncircular orbital motion (see~\PaperI).

\section{Discussion}\label{sec:discussion}

This paper demonstrates, for the first time, practical lattice template placement for an all-sky broadband-frequency search for gravitational-wave pulsars.
The mismatch predictions of the reduced supersky metric, derived previously in~\PaperI, are confirmed using realistic mismatch distributions, as would be encountered in a real search.
An algorithm implementing lattice template placement, including efficient iteration over the parameter space, nearest template finding, and correct treatment of the parameter-space boundaries, is described and tested.
The number of templates generated by the algorithm is consistent with theoretical expectations, and with previous results in the literature.

Future work will extend the reduced supersky metric to a semicoherent search.
In such a search, the data are partitioned into $N$ data segments, and an average $\calF$-statistic is computed on a high-resolution ``fine'' template bank from $N$ $\calF$-statistic values, which are computed by coherently match-filtering each segment on $N$ low-resolution ``coarse'' template banks.
Typically, the template in each coarse template bank which is closest to the current fine template is chosen to contribute to the average $\calF$-statistic; this requires an algorithm for finding the nearest template, as that described in Section~\ref{sec:near-templ-find-idx}.
The resolution of the fine template bank is determined by the parameter-space metric of the average $\calF$-statistic, which is the average of the parameter-space metrics of each data segment~\cite{Brady.Creighton.2000a}.
The extension of the reduced supersky metric to an averaged metric will need to be investigated.

While the focus of this paper has been searches for isolated gravitational-wave pulsars, the lattice template algorithm presented here may be applied to any parameter-space described by a constant metric.
An interesting example is searches for gravitational-wave pulsars in low-mass X-ray binary systems with a known sky location~\cite{Messenger.2011a}, where the parameter space comprises frequency and the orbital parameters of the binary system.
Recent work on the parameter-space metric~\cite{Leaci.Prix.2014} focuses on the feasibility of a search for Scorpius X-1, one of the most promising gravitational-wave sources of this type.

\acknowledgments

I thank Reinhard Prix for many valuable discussions.
Numerical simulations were performed on the ATLAS computer cluster of the Max-Planck-Institut f\"ur Gravitationsphysik.
This paper has document numbers AEI-2014-052 and LIGO-P1400202.

\appendix

\onecolumngrid

\section{Approximations to the reduced supersky metric sky elements}

The elements $g\unana$ and $g\unbnb$ of the reduced supersky metric $\mat g\urss$ are approximated by the following expressions, where $\log T \equiv \log(T / 1~\mathrm{s})$: for first spindown,
\begin{subequations}
\label{eqs:metric-fit-gnabnab-eqn}
\begin{align}
\log\frac{g\unana}{f\umax^2} &\approx \begin{cases}
\frac{(\log T)^2}{3.803} - \frac{\log T}{0.2836} \,, & 0.5 \le \frac{T}{\text{day}} < 1.38 \,,\\
\frac{(\log T)^2}{19.04} - \frac{\log T}{0.9374} \,, & 1.38 \le \frac{T}{\text{day}} < 10.5 \,,\\
\frac{(\log T)^2}{2.669} - \frac{\log T}{0.1823} \,, & 10.5 \le \frac{T}{\text{day}} < 60 \,,
\end{cases} & \log\frac{g\unbnb}{f\umax^2} &\approx \begin{cases}
\frac{(\log T)^2}{7.751} - \frac{\log T}{0.4736} \,, & 0.5 \le \frac{T}{\text{day}} < 0.875 \,,\\
\frac{(\log T)^2}{3.245} - \frac{\log T}{0.2425} \,, & 0.875 \le \frac{T}{\text{day}} < 1.75 \,,\\
\frac{(\log T)^2}{24.81} - \frac{\log T}{1.076} \,, & 1.75 \le \frac{T}{\text{day}} < 29 \,,\\
\frac{(\log T)^2}{2.5} - \frac{\log T}{0.1605} \,, & 29 \le \frac{T}{\text{day}} < 60 \,;
\end{cases}
\end{align}
and for second spindown,
\begin{align}
\log\frac{g\unana}{f\umax^2} &\approx \begin{cases}
\frac{(\log T)^2}{7.629} - \frac{\log T}{0.4687} \,, & 0.5 \le \frac{T}{\text{day}} < 0.875 \,,\\
\frac{(\log T)^2}{3.45} - \frac{\log T}{0.2553} \,, & 0.875 \le \frac{T}{\text{day}} < 1.88 \,,\\
\frac{(\log T)^2}{26.62} - \frac{\log T}{1.123} \,, & 1.88 \le \frac{T}{\text{day}} < 28 \,,\\
\frac{(\log T)^2}{2.591} - \frac{\log T}{0.1663} \,, & 28 \le \frac{T}{\text{day}} < 60 \,,
\end{cases} & \log\frac{g\unbnb}{f\umax^2} &\approx \begin{cases}
\frac{(\log T)^2}{12.64} - \frac{\log T}{0.6331} \,, & 0.5 \le \frac{T}{\text{day}} < 1 \,,\\
\frac{(\log T)^2}{3.194} - \frac{\log T}{0.2359} \,, & 1 \le \frac{T}{\text{day}} < 2.12 \,,\\
\frac{(\log T)^2}{22.67} - \frac{\log T}{1.021} \,, & 2.12 \le \frac{T}{\text{day}} < 60 \,.
\end{cases}
\end{align}
\end{subequations}
These are derived by numerically computing $\mat g\urss$ at a fixed $f\umax$, and over the following ranges of time span $T$ and start time $t\ustart$: $T$ from 0.5~to 3~days in steps of 0.125~days, then from 3~to 60~days in steps of 0.5~days; $t\ustart = t\uref + \Delta t\uref - 0.5 T$, where $t\uref$ is one of UTC~2000-06-18 23:59:47, 2009-01-05 11:59:45, or 2017-07-24 23:59:44, and $\Delta t\uref$ from 0~to 720~days in steps of 1~day.
Sinusoidal modulations in $g\unana$ and $g\unbnb$ with respect to $t\ustart$, with periods of a synodic month and a year, were $\lesssim 10\%$.
The two expressions $\log (\langle g_{\sigma \sigma} \rangle / f\umax^2) / \log T$, where $\sigma \in \{ n\ua, n\ub \}$ and $\langle\cdot\rangle$ denotes averaging over $t\ustart$, are then fitted by piecewise linear functions in $\log T$.

\twocolumngrid

\bibliography{paper}

\end{document}